\documentclass[sigconf]{acmart}

\usepackage{subfigure}
\usepackage{graphicx}
\usepackage{amsmath}
\usepackage{bm}
\usepackage{url}
\usepackage{stmaryrd}
\usepackage{balance}
\usepackage{amssymb}
\usepackage{pgfplots}
\usepackage{pifont}
\usepackage{epsfig}
\usepackage{booktabs}
\usepackage{pgffor}
\usepackage{tikz}
\usepackage{multirow}
\usepackage{amsmath}
\usepackage{tikz}
\usetikzlibrary{positioning,arrows,calc}
\usepackage[all]{xy}
\usepackage{makecell}
\usepackage{enumitem}
\usepackage{textcomp,libertine}

\makeatletter
\newcommand\xleftrightarrow[2][]{%
	\ext@arrow 9999{\longleftrightarrowfill@}{#1}{#2}}
\newcommand\longleftrightarrowfill@{%
	\arrowfill@\leftarrow\relbar\rightarrow}
\makeatother

\newcommand{\mb}{\mathbf}

\newcommand{\Rho}{\mathrm{P}}

\newcommand{\our}{\textsl{SHNA}}
\newcommand{\sppu}{\textsl{SpectralIter}}
\newcommand{\kmeanspu}{\textsl{KmeansIter}}
\newcommand{\pusvm}{\textsl{IterClip-MD}}
\newcommand{\svm}{\textsl{Mna-MD}}
\newcommand{\dw}{\textsl{DeepWalk}}
\newcommand{\mv}{\textsl{Metapath2vec}}

\usepackage{algorithm}
\usepackage{algorithmic}

\begin{document}
\title{Scalable Heterogeneous Social Network Alignment through Synergistic Graph Partition}

\author{Yuxiang~Ren}
\affiliation{%
	\institution{IFM Lab\\ Department of Computer Science\\ Florida State University, FL, USA\\}
}
\email{yuxiang@ifmlab.org}
\author{Lin~Meng}
\affiliation{%
	\institution{IFM Lab\\ Department of Computer Science\\ Florida State University, FL, USA\\}
}
\email{lin@ifmlab.org}
\author{Jiawei~Zhang}
\affiliation{%
	\institution{IFM Lab\\ Department of Computer Science\\ Florida State University, FL, USA\\}
}
\email{jiawei@ifmlab.org}

\begin{abstract}
	Social network alignment has been an important research problem for social network analysis in recent years. With the identified shared users across networks, it will provide researchers with the opportunity to achieve a more comprehensive understanding of users' social activities both within and across networks. Social network alignment is a very difficult problem. 
	Besides the challenges introduced by the network heterogeneity, the network alignment can be reduced to a combinatorial optimization problem with an extremely large search space. The learning effectiveness and efficiency of existing alignment models will be degraded significantly as the network size increases. In this paper, we focus on studying the scalable heterogeneous social network alignment problem, and propose to address it with a novel two-stage network alignment model, namely  \textbf{S}calable \textbf{H}eterogeneous \textbf{N}etwork \textbf{A}lignment ({\our}). Based on a group of intra- and inter-network meta diagrams, {\our} first partitions the social networks into a group of sub-networks synergistically. Via the partially known anchor links, {\our} can extract the partitioned sub-network correspondence relationships. Instead of aligning the complete input network, {\our} proposes to identify the anchor links between the matched sub-network pairs, while those between the unmatched sub-networks will be pruned to effectively shrink the search space. Extensive experiments have been done to compare {\our} with the state-of-the-art baseline methods on a real-world aligned social networks dataset. The experimental results have demonstrated both the effectiveness and efficiency of {\our} in addressing the problem.
	
\end{abstract}
\keywords{Heterogeneous Network; Network Alignment; Synergistic Partition}

\maketitle

\section{Introduction}\label{sec:introduction}

In recent years, a large number of online social networks have appeared, which can provide people with various kinds of services. To enjoy these different services at the same time, users nowadays are usually involved in a number of online social networks simultaneously. For instance, people will join in Facebook\footnote{https://www.facebook.com} to socialize with their friends; use Linkedin\footnote{https://www.linkedin.com} to establish their professional profile; rely on Twitter\footnote{https://twitter.com} to access and comment on the latest news information. However, in the real world, these different online social networks are mostly isolated without any knowledge about the shared users among them, which renders the inter-network social network analysis a great challenge.

Recently, some research works have proposed to study the alignment problem \cite{KZY13,ZP19} across multiple online social networks. The main objective of the social network alignment problem is to uncover the mappings of common users across networks, which are named as the \textit{anchor links}  \cite{KZY13} formally. 
Social network alignment provides researchers with the opportunity to study the users' social activities from a global perspective. By integrating the social activity information from multiple social sites, we can achieve a more comprehensive knowledge about users' social preferences. Meanwhile, via these inferred anchor links, information can also propagate across different social networks to improve the services of different social networks simultaneously. 


Formally, given two networks $G^{(1)}$ and $G^{(2)}$ with $m$ and $n$ users respectively, we can denote the number of true anchor links existing between $G^{(1)}$ and $G^{(2)}$ as $l$. According to \cite{KZY13}, the anchor links to be inferred are usually subject to the \textit{one-to-one} cardinality constraint. In other words, each user will be connected by at most one anchor link between the networks, and we can have $l \le \min(m, n)$. Social network alignment problem aims at identifying these $l$ true anchor links from the $m \times n$ potential anchor links across networks $G^{(1)}$ and $G^{(2)}$, which will lead to a combinatorial optimization problem of time complexity $O\left(\dbinom{m \times n}{\min(m, n)} \right)$. Most of the existing network alignment models are mainly proposed based on the complete input network \cite{KZY13,ZY15_ijcai,ZCZCY17}, which will become ineffective for large-scale online social networks with a large number of users.

\noindent \textbf{Problem Studied}: In this paper, we will study the scalable online social network alignment problem, where each social network studied is of a heterogeneous structure involving multiple types of nodes and links. To address the problem, a reduction of the search space, i.e., these aforementioned $m \times n$ potential anchor links, is necessary and critical, which can not only improve the learning effectiveness but also significantly lower down the time costs in model learning.


The heterogeneous social network alignment problem is extremely challenging to address due to several different reasons:
\begin{itemize}
	\item \textit{Heterogeneity}: There exist various types of heterogeneous information in the online social networks, which can provide critical signals for identifying the common users across networks. Meanwhile, properly handling such heterogeneous information in a unified way is not an easy task. 
	
	\item \textit{Scalability}: For the large-sized input online social networks, besides the effectiveness, learning efficiency is another crucial factor to consider in the model building. Few of the existing research works have ever studied this problem yet, which remains an open problem by this context so far.
	
	\item \textit{Generalizability}: To ensure the applicability of the proposed model
	, we need to propose a general learning model that can be extensible to various learning settings. Besides differentiating the non-existing anchor links from the real ones, the model should also incorporate the \textit{one-to-one} cardinality constraint \cite{KZY13} into the learning process effectively.
	
\end{itemize}


To address these challenges aforementioned, we introduce a novel scalable heterogeneous social network alignment framework, namely \textbf{S}calable \textbf{H}eterogeneous \textbf{N}etwork \textbf{A}lignment ({\our}), in this paper. To effectively capture the diverse connections among users within and across networks with heterogeneous information, {\our} employs a group of \textit{meta diagrams} in this paper. \textit{Meta diagram} is a novel concept proposed in \cite{RAZ19}, which includes both \textit{meta path} and more complex meta structures to outline the user correlations both within and across heterogeneous networks
. As a scalable and general solution, {\our} addresses the social network alignment problem via two stages: \textit{network synergistic partition} and \textit{parallel sub-network alignment}. {\our} proposes to partition the large-sized input social network data into a group sub-networks with a synergistic network partition method
. The partition process needs to take care of both the diverse intra- and inter-network user connections, where the shared users should be partitioned into the groups with correspondence relationships as indicated by the partially known anchor links. Then alignment will be performed between these identified corresponding sub-networks only in the second stage, whose learning results will be fused to recover the complete alignment result of the input networks.


The remaining parts of this paper are organized as follows. In Section~\ref{sec:formulation}, we introduce the definitions of several important terminologies and the formal problem statement. Meta diagram which is the basis of features in this paper is introduced in Section~\ref{sec:feature}. Detailed information about the proposed model is provided in Section~\ref{sec:method}, whose effectiveness and efficiency are verified in Section~\ref{sec:experiment}. Related works are discussed in Section~\ref{sec:related_work} and finally in Section~\ref{sec:conclusion} we conclude this paper.

\section{Problem Formulation} \label{sec:formulation}
\subsection{Terminology Definition}
The network we study is an \textit{attributed heterogeneous social network}. 
\begin{figure}[t]
	\centering
	\begin{minipage}[l]{1.0\columnwidth}
		\centering
		\includegraphics[width=\textwidth]{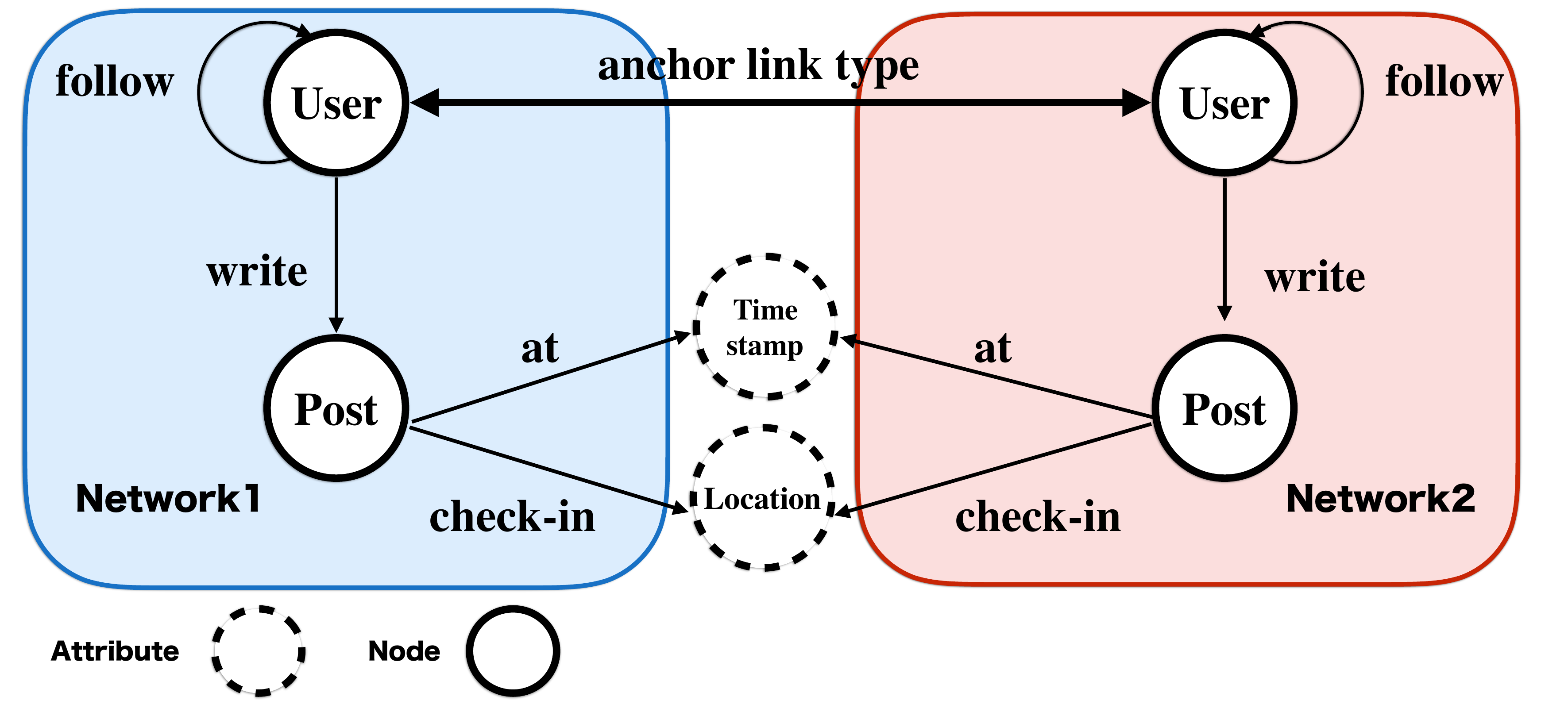}
	\end{minipage}
	\vspace{-5pt}
	\caption{Schema of aligned networks.}\label{fig:schema}\vspace{-15pt}
\end{figure}
\noindent \textbf{Definition 1}
\noindent(Attributed Heterogeneous Social Network): The \textit{attributed heterogeneous social network} can be represented as $G = (\mathcal{V}, \mathcal{E}, \mathcal{T})$. $\mathcal{V} = \bigcup_i \mathcal{V}_i$ is the set of different nodes, while $\mathcal{E} = \bigcup_i \mathcal{E}_i$ represents the set of complex links in the network. Besides, the set $\mathcal{T}$ = $\bigcup_i \mathcal{T}_i$ represents a group of attributes attached to the nodes.

Among multiple \textit{attributed heterogeneous social networks}, if there exist shared users, we can define them as \textit{aligned attributed heterogeneous social networks}

\noindent \textbf{Definition 2}
(Aligned Attributed Heterogeneous Social Networks): Given the \textit{attributed heterogeneous social networks} $G^{(1)}$, $G^{(2)}$ and common users are shared between them, we can define them as the \textit{aligned attributed heterogeneous social networks} $\mathcal{G} = \left( (G^{(1)}, G^{(2)}), \mathcal{A}^{(1,2)} \right)$, and $\mathcal{A}^{(1,2)}$ is the set of undirected anchor links between $G^{(1)}$ and $G^{(2)}$ which connect the common users.

Here, we take two famous online social networks Foursquare and Twitter as an example. We represent them as $\mathcal{G} = ((G^{(1)}, G^{(2)}),\mathcal{A}^{(1,2)})$, where $G^{(1)}$ represents Foursquare and $G^{(2)}$ is Twitter. The Foursquare network $G^{(1)}$ can be represented as $G^{(1)} = (\mathcal{V}^{(1)}, \mathcal{E}^{(1)}, \mathcal{T}^{(1)})$, where $\mathcal{V}^{(1)}$ is the union of $\mathcal{U}^{(1)}$ and ${Post}^{(1)}$ representing the sets of users and posts in the network respectively. $\mathcal{E}^{(1)} = \mathcal{E}_{u,u}^{(1)} \cup \mathcal{E}_{u,p}^{(1)}$ contains the set of social links among users and the set of write links between users and posts. $\mathcal{T}^{(1)} = \mathcal{T}^{(1)}_{l} \cup \mathcal{T}^{(1)}_{t}$ denotes the set of attributes extracted from the posts in ${Post}^{(1)}$ including location checkins $\mathcal{T}^{(1)}_{l}$ and timestamps $\mathcal{T}^{(1)}_{t}$ in this example. The Twitter network can be represented in a similar format as Foursquare, which can be denoted as $G^{(2)} = (\mathcal{V}^{(2)}, \mathcal{E}^{(2)}, \mathcal{T}^{(2)})$. User anchor links in set $\mathcal{A}^{(1,2)}$ connecting to shared users between Foursquare and Twitter can effectively align these two networks together. In the following parts, we illustrate the problem setting and the proposed framework based on the aligned Foursquare and Twitter networks, i.e., $\mathcal{G}$.
\subsection{Problem Definition}

\noindent Given \textit{aligned attributed heterogeneous social networks} $\mathcal{G}$, we can represent all potential anchor links between networks $G^{(1)}$ and $G^{(2)}$ as set $\mathcal{H} = \mathcal{U}^{(1)} \times \mathcal{U}^{(2)}$, where $\mathcal{U}^{(1)}$ and $\mathcal{U}^{(2)}$ denote the user sets in $G^{(1)}$ and $G^{(2)}$ respectively. For the known anchor links, we can group them as a labeled set  $\mathcal{D} = \mathcal{A}^{(1,2)}$. The remaining anchor links with unknown labels are those to be inferred, and they can be denoted as the unlabeled set $\mathcal{P} = \mathcal{H} \setminus \mathcal{D} $. Based on both $\mathcal{D}$ and $\mathcal{P}$, we aim at building a mapping function $f: \mathcal{H} \to \mathcal{Y}$ to infer anchor link labels in $\mathcal{Y} = \{0, +1\}$ subject to the \textit{one-to-one} constraint, where labels $+1$ and $0$ denote the existing and non-existing anchor links respectively. 

\begin{figure*}[t]
	\centering
	\begin{minipage}[l]{1.8\columnwidth}
		\centering
		\includegraphics[width=\textwidth]{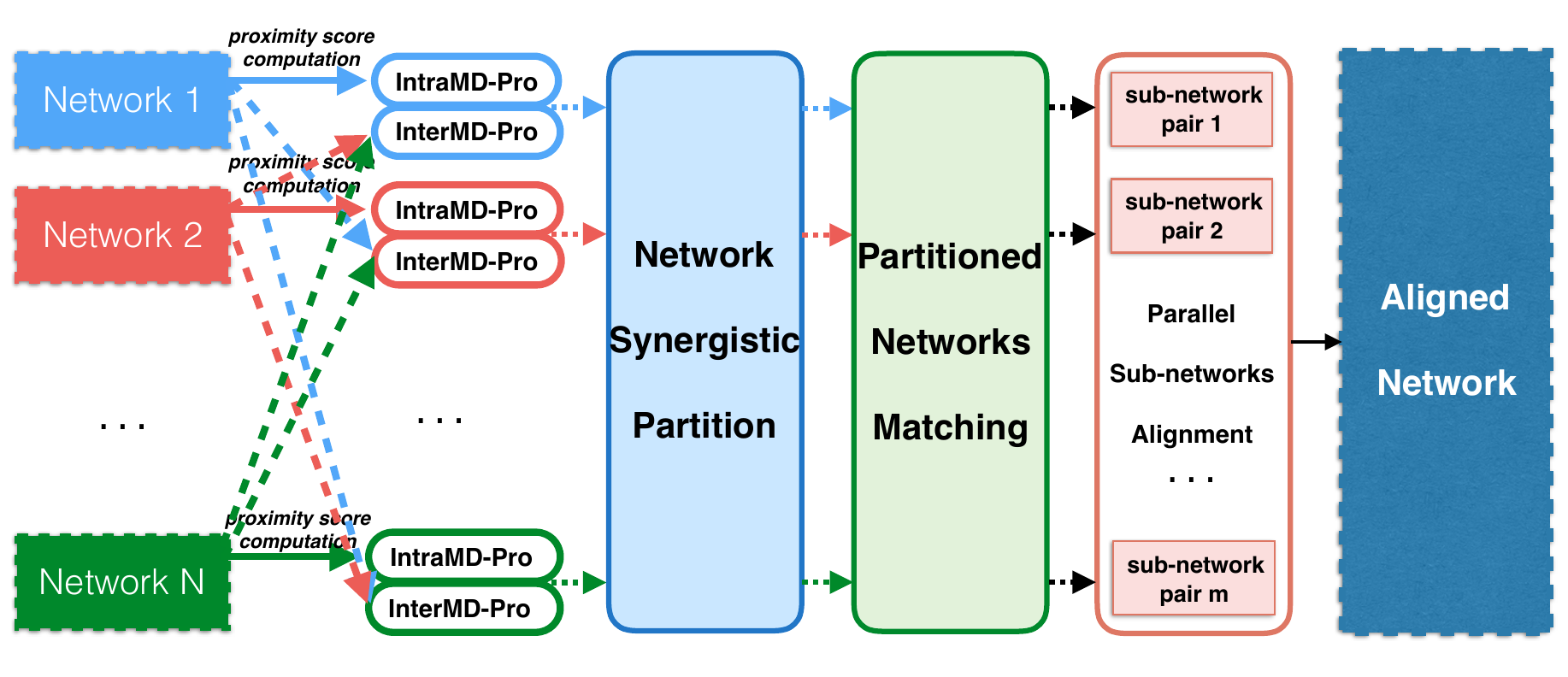}
	\end{minipage}
	\vspace{-5pt}
	\caption{Parallel Implementation of Framework {\our}}\label{fig:framework}
	\vspace{-15pt}
\end{figure*}
\vspace{-2pt}
\section{Meta Diagram}\label{sec:feature}
Before introducing the {\our} framework, we first introduce \textit{intra-network meta diagram} and \textit{inter-network meta diagram}, which will be used to measure the proximity among users in the {\our}.

\subsection{Network Schema}
In order to better understand the complex \textit{aligned attributed heterogeneous social networks}, it is necessary to define the schema-level description. 
\noindent \textbf{Definition 3} (Aligned Attributed Heterogeneous Social Network Schema): 
The schema of the given aligned social networks $\mathcal{G}$ can be represented as $S_{\mathcal{G}} = ((S_{{G}^{(1)}}, S_{{G}^{(2)}}),\{\mbox{anchor}\})$. Here, $S_{{G}^{(1)}} = (\mathcal{N}^{(1)}_{\mathcal{V}} \cup \mathcal{N}^{(1)}_{\mathcal{T}}, \mathcal{R}^{(1)}_{\mathcal{E}} \cup \mathcal{R}^{(1)}_{\mathcal{A}})$, where $\mathcal{N}^{(1)}_{\mathcal{V}}$ and $\mathcal{N}^{(1)}_{\mathcal{T}}$ denote the set of node types and attribute types in the network, while  $\mathcal{R}^{(1)}_{\mathcal{E}}$ represents the set of link types in the network, and $\mathcal{R}^{(1)}_{\mathcal{A}}$ denotes the set of association types between nodes and attributes. In a similar way, the schema of $G^{(2)}$ is $S_{{G}^{(2)}} = (\mathcal{N}^{(2)}_{\mathcal{V}} \cup \mathcal{N}^{(2)}_{\mathcal{T}}, \mathcal{R}^{(2)}_{\mathcal{E}} \cup \mathcal{R}^{(2)}_{\mathcal{A}})$.
We display the schema of the Foursquare and Twitter networks in Figure~\ref{fig:schema}, where the exact node, attribute, and link types can be found intuitively.

\begin{table*}[t]

	\scriptsize
	\centering
	{
		\caption{Summary of Inter-Network Meta Diagrams.}\label{tab:inter_meta_diagram}
		\vspace{-5pt}
		\begin{tabular}{llll}
			\hline
			\textbf{ID}
			&\textbf{Notation}
			& \textbf{Meta Diagram}
			& \textbf{Semantics}\\
			\hline
			
			$\Rho^{A}_1$
			&U $\to$ U $\leftrightarrow$ U $\gets$ U
			&User $\xrightarrow{follow}$ User $\xleftrightarrow{anchor}$ User $\xleftarrow{follow}$ User
			&Common Anchored Followee\\
			
			$\Rho^{A}_2$
			&U $\gets$ U $\leftrightarrow$ U $\to$ U
			&User $\xleftarrow{follow}$ User $\xleftrightarrow{anchor}$ User $\xrightarrow{follow}$ User
			&Common Anchored Follower\\
			
			$\Rho^{A}_3$
			&U $\to$ U $\leftrightarrow$ U $\to$ U
			&User $\xrightarrow{follow}$ User $\xleftrightarrow{anchor}$ User $\xrightarrow{follow}$ User
			&Common Anchored Followee-Follower\\
			
			$\Rho^{A}_4$
			&U $\gets$ U $\leftrightarrow$ U $\gets$ U
			&User $\xleftarrow{follow}$ User $\xleftrightarrow{anchor}$ User $\xleftarrow{follow}$ User
			&Common Anchored Follower-Followee\\

			$\Rho^{A}_5$
			&U $\to$ P $\to$ T $\gets$ P $\gets$ U
			&User $\xrightarrow{write}$ Post $\xrightarrow{at}$ Timestamp $\xleftarrow{at}$ Post $\xleftarrow{write}$ User
			&Common Timestamp\\
			
			$\Rho^{A}_6$
			&U $\to$ P $\to$ L $\gets$ P $\gets$ U
			&User $\xrightarrow{write}$ Post $\xrightarrow{checkin}$ Location $\xleftarrow{checkin}$ Post $\xleftarrow{write}$ User
			&Common Checkin\\
			
			\hline
			
			$\Psi^{A}_1 (\Rho^{A}_1 \times \Rho^{A}_2)$
			
			&U $\leftrightarrow$ U $\xleftrightarrow{anchor}$ U $\leftrightarrow$ U
			
			&
			\begin{tikzpicture}[baseline={([yshift=-.5ex]current bounding box.center)}, node distance=1cm, auto,]
			\node[] (dummy) {};
			\coordinate (CENTER) at ($(dummy)$);
			
			\node[right=of dummy] (post-right) {User};
			
			\node[left=of dummy] (post-left) {User}
			edge[<->] node[midway,above] {{\tiny $anchor$}} (post-right.west);

			\node[above=0.01cm of post-right] (user-right-above) {};
			\node[above=0.01cm of post-left] (user-left-above) {};
			
			\node[below=0.01cm of post-right] (user-right-below) {};
			\node[below=0.01cm of post-left] (user-left-below) {};
			
			\node[left=of post-left] (user-left) {User}
			edge[->, bend left=10] node[midway,above] {{\tiny $follow$}} (user-left-above.west)
			edge[<-, bend left=-10] node[midway,below] {{\tiny $follow$}} (user-left-below.west);
			
			\node[right=of post-right] (user-right) {User}
			edge[->, bend left=-10] node[midway,above] {{\tiny $follow$}} (user-right-above.east)
			edge[<-, bend left=10] node[midway,below] {{\tiny $follow$}} (user-right-below.east);
			
			\end{tikzpicture}
			&\makecell[l]{Common Aligned Neighbors}\\

			$\Psi^{A}_2 (\Rho^{A}_5 \times \Rho^{A}_6)$
			&
			\begin{tikzpicture}[baseline={([yshift=-.5ex]current bounding box.center)}, node distance=1cm]
			\node (U) at (-.15, 0) {U};
			\node (P) at (0.5, 0) {P};
			\node (T) at (1.15, -0.3) {T};
			\node (L) at (1.15, 0.3) {L};
			\node (P2) at (1.8, 0) {P};
			\node (U2) at (2.45, 0) {U};
			
			\draw[->, to path={|- (\tikztotarget)}]
			(U) edge (P) (P) edge (L) (P) edge (T);
			\draw[<-, to path={-| (\tikztotarget)}]
			(L) edge (P2) (T) edge (P2);
			\draw[<-, to path={-| (\tikztotarget)}]
			(P2) edge (U2.west);
			
			\end{tikzpicture}&User $\xrightarrow{write}$  
			\begin{tikzpicture}[baseline={([yshift=-.5ex]current bounding box.center)}, node distance=1cm, auto,]
			\node[] (dummy) {};
			\coordinate (CENTER) at ($(dummy)$);
			\node[inner sep=0pt,above=0.0cm of dummy] (location) {Location};
			\node[inner sep=0pt,below=0.0cm of dummy] (time) {Timestamp};
			
			\node[right=of dummy] (post-right) {Post}
			edge[->, bend right=10] node[midway,above] {{\tiny $checkin$}} (location.east) 
			edge[->, bend left=10] node[midway,below] {{\tiny $at$}} (time.east); 
			
			\node[left=of dummy] (post-left) {Post}
			edge[->, bend left=10] node[midway,above] {{\tiny $checkin$}} (location.west)
			edge[->, bend right=10] node[midway,below] {{\tiny $at$}} (time.west);
			\end{tikzpicture}
			$\xleftarrow{write}$ User
			&\makecell[l]{Common Attributes}\\
			
			$\Psi^{A}_3 (\Rho^{A}_1 \times \Rho^{A}_5 \times \Rho^{A}_6)$
			&
			\begin{tikzpicture}[baseline={([yshift=-.5ex]current bounding box.center)}, node distance=1cm]
			\node (U11) at (-.15, 0.5) {U};
			\node (U) at (-.15, 0) {U};
			\node (P) at (0.5, 0) {P};
			\node (W) at (1.15, -0.2) {T};
			\node (L) at (1.15, 0.2) {L};
			\node (P2) at (1.8, 0) {P};
			\node (U2) at (2.45, 0) {U};
			\node (U22) at (2.45, 0.5) {U};
			
			\draw[->, to path={|- (\tikztotarget)}]
			(U) edge (P) (P) edge (L) (P) edge (W) (U) edge (U11.south) (U2) edge (U22.south);
			\draw[<-, to path={-| (\tikztotarget)}]
			(L) edge (P2) (W) edge (P2);
			\draw[<-, to path={-| (\tikztotarget)}]
			(P2) edge (U2.west);
			\draw[<-, to path={-| (\tikztotarget)}]
			(U11) edge (U22.west);
			\draw[->, to path={|- (\tikztotarget)}]
			(U11) edge (U22.west);
			\end{tikzpicture}
			
			&
			\begin{tikzpicture}[baseline={([yshift=-.5ex]current bounding box.center)}, node distance=1cm, auto,]
			\node[] (dummy) {};
			\coordinate (CENTER) at ($(dummy)$);
			\node[inner sep=0pt,above=0.0cm of dummy] (location) {Location};
			\node[inner sep=0pt,below=0.0cm of dummy] (time) {Timestamp};
			
			\node[left=of dummy] (post-left) {Post}
			edge[->, bend left=10] node[midway,above] {{\tiny $checkin$}} (location.west)
			edge[->, bend right=10] node[midway,below] {{\tiny $at$}} (time.west);
			
			\node[right=of dummy] (post-right) {Post}
			edge[->, bend right=10] node[midway,above] {{\tiny $checkin$}} (location.east) 
			edge[->, bend left=10] node[midway,below] {{\tiny $at$}} (time.east); 
			
			\node[above=0.2cm of post-right] (user-right-above) {User};
			\node[above=0.2cm of post-left] (user-left-above) {User}
			edge[<->] node[midway,above] {{\tiny $anchor$}} (user-right-above.west);
			
			\node[left=of post-left] (user-left) {User}
			edge[->] node[midway,above] {{\tiny $write$}} (post-left.west)
			edge[->, bend left=10] node[midway,above] {{\tiny $follow$}} (user-left-above.west);
			
			\node[right=of post-right] (user-right) {User}
			edge[->] node[midway,above] {{\tiny $write$}} (post-right.east)
			edge[->, bend left=-10] node[midway,above] {{\tiny $follow$}} (user-right-above.east);
			
			\end{tikzpicture}
			&\makecell[l]{Common Aligned Neighbor \& Attributes}\\

			\hline
			
		\end{tabular}
	}
\end{table*}

\subsection{Inter-Network Meta Diagram}\label{sec:inter}
The definition of inter-network meta diagram is first proposed in \cite{RAZ19}. Based on our own problem, the definition of \textit{inter-network meta diagram} can be presented as follows:

\noindent \textbf{Definition 4}
(Inter-Network Meta Diagram): Given \textit{aligned attributed heterogeneous social networks} $S_{\mathcal{G}} = ((S_{{G}^{(1)}}, S_{{G}^{(2)}}), \{\mbox{anchor}\})$. An \textit{inter-network meta diagram} can be formally represented as a directed acyclic subgraph $\Psi^{A} = (\mathcal{N}_{\Psi}, \mathcal{R}_{\Psi}, N_s, N_t)$, where $\mathcal{N}_{\Psi} \subset (\mathcal{N}^{(1)}_{\mathcal{V}} \cup \mathcal{N}^{(2)}_{\mathcal{V}}  \cup \mathcal{N}^{(1)}_{\mathcal{T}}  \cup \mathcal{N}^{(2)}_{\mathcal{T}})$ and $\mathcal{R}_{\Psi} \subset (\mathcal{R}^{(1)}_{\mathcal{E}} \cup \mathcal{R}^{(2)}_{\mathcal{E}} \cup \mathcal{R}^{(1)}_{\mathcal{A}} \cup \mathcal{R}^{(2)}_{\mathcal{A}} \cup \{\mbox{anchor}\})$. $N_s, N_t $ denote the source and target node types from $G^{(1)}$ and $G^{(2)}$ respectively.

The notaion, description and physical meanings of \textit{inter-network meta paths} used in this paper are summarized in the first section of Table~\ref{tab:inter_meta_diagram}. Because of the problem we try to solve, we are concerned about \textit{inter-network meta diagrams} connecting two users from different networks. We list several \textit{inter-network meta diagram} examples in the second section of Table~\ref{tab:inter_meta_diagram} which can be represented as $\{\Psi^{A}_1, \Psi^{A}_2, \Psi^{A}_3\}$. Now we focus on the $\Psi^{A}_1$ at first. It is composed of two meta paths which are both $\Rho^{A}_1$ and represent two users have two followees respectively where there exits an anchor link between these two followees. $\Psi^{A}_2$ is built by $\Rho^{A}_5$ and $\Rho^{A}_6$ which represents two users have posts checking in the same location and at the same time. $\Psi^{A}_3$ containing 3 inter-network meta paths $\Rho^{A}_1$, $\Rho^{A}_5$ and $\Rho^{A}_6$. In a more formal way, we can classify inter-network meta paths as $\Rho^{A}_{f}$ containing the social relationship based inter-network meta paths and $\Rho^{A}_{a}$ representing the sets of the attribute based paths, where $\Rho^{A}_{f} = \{\Rho^{A}_1, \Rho^{A}_2, \Rho^{A}_3, \Rho^{A}_4\}$ and $\Rho^{A}_{a} = \{\Rho^{A}_5, \Rho^{A}_6\}$. Besides, we also define that $\Rho^{A} = \Rho^{A}_{a} \cup \Rho^{A}_{f} $. Therefore, we can list inter-network meta diagrams used in {\our} in Table~\ref{tab:inter_stack}. We can represent inter-network meta diagrams as $\Psi^{A} = \Rho^{A} \cup \Psi^{A}_{f^2} \cup \Psi^{A}_{a^2} \cup \Psi^{A}_{f,a} \cup \Psi^{A}_{f,a^2} \cup \Psi^{A}_{f^2,a^2}$.
\begin{table}[h]
	\caption{Inter-network Meta Diagrams}
	\label{tab:inter_stack}
	\vspace{-8pt}
	\begin{tabular}{p{25mm}|cl}
		\toprule
		Set &Physical Meanings \\
		\midrule
		\scriptsize$\Psi^{A}_{f^2}$ ($\Rho^{A}_{f} \times \Rho^{A}_{f}$) & Common Aligned Neighbor\underline{\textbf{s}}  \\
		\midrule
		\scriptsize$\Psi^{A}_{a^2}$ ($\Rho^{A}_{a} \times \Rho^{A}_{a}$) & Common Attribute\underline{\textbf{s}}\\
		\midrule
		\scriptsize$\Psi^{A}_{f,a}$ ($\Rho^{A}_{f} \times \Rho^{A}_{a}$) & Common Aligned Neighbor \& Attribute\\
		\midrule
		\scriptsize$\Psi^{A}_{f,a^2}$ ($\Rho^{A}_{f} \times \Rho^{A}_{a} \times \Rho^{A}_{a}$) & Common Aligned Neighbor \& Attribute\underline{\textbf{s}}\\
		\midrule
		\scriptsize$\Psi^{A}_{f^2,a^2}$ ($\Rho^{A}_{f} \times \Rho^{A}_{f} \times \Rho^{A}_{a} \times \Rho^{A}_{a}$) & Common Aligned Neighbor\underline{\textbf{s}} \& Attribute\underline{\textbf{s}}\\
		\bottomrule
	\end{tabular}
	\vspace{-15pt}
\end{table}

\begin{table*}[t]
	\scriptsize
	\centering
	{
		\caption{Summary of Intra-Network Meta Diagram.}\label{tab:intra_meta_diagram}
		\vspace{-5pt}
		\begin{tabular}{llll}
			\hline
			\textbf{ID}
			&\textbf{Notation}
			& \textbf{Meta Diagram}
			& \textbf{Semantics}\\
			\hline
			\hline
			
			$\Rho^{I}_1$
			&U $\to$ U
			&User $\xrightarrow{follow}$ User
			&Follow\\
			
			$\Rho^{I}_2$
			&U $\to$ U $\to$ U
			&User $\xrightarrow{follow}$ User $\xrightarrow{follow}$ User
			&Follower of Follower\\
			
			$\Rho^{I}_3$
			&U $\to$ U $\gets$ U
			&User $\xrightarrow{follow}$ User $\xrightarrow{follow^{-1}}$ User
			&Common Out Neighbor\\
			
			$\Rho^{I}_4$
			&U $\gets$ U $\to$ U
			&User $\xrightarrow{follow^{-1}}$ User $\xrightarrow{follow}$ User
			&Common In Neighbor\\

			$\Rho^{I}_5$
			&U $\to$ P $\to$ T $\gets$ P $\gets$ U
			&User $\xrightarrow{write}$ Post $\xrightarrow{at}$ Timestamp $\xleftarrow{at}$ Post $\xleftarrow{write}$ User
			&Posts Containing Common Timestamps\\

			$\Rho^{I}_6$
			&U $\to$ P $\to$ L $\gets$ P $\gets$ U
			&User $\xrightarrow{write}$ Post $\xrightarrow{checkin}$ Location $\xleftarrow{checkin}$ Post $\xleftarrow{write}$ User
			&Posts Attaching Common Location Check-ins\\
			
			\hline
			
			$\Psi^{I}_1 (\Rho^{I}_1 \times \Rho^{I}_1)$
			
			&U $\xleftrightarrow{}$ U
			
			&
			\begin{tikzpicture}[baseline={([yshift=-.5ex]current bounding box.center)}, node distance=1cm, auto,]
			\node[] (dummy) {};
			\coordinate (CENTER) at ($(dummy)$);
			
			\node[left=of dummy] (post-left) {User};
			\node[left=of post-left] (left) {User};
			

			\node[above=0.0001cm of post-left] (post-user-left-above) {};
			\node[below=0.0001cm of post-left] (post-user-left-below) {};
			\node[above=0.0001cm of left] (user-left-above) {}
			edge[->, bend left=5] node[midway,above] {{\tiny $follow$}} (post-user-left-above);
			
			\node[below=0.0001cm of left] (user-left-below) {}
			edge[<-, bend left=-5] node[midway,below] {{\tiny $follow$}} (post-user-left-below);

			\end{tikzpicture}
			&\makecell[l]{Follower and Followee}\\

			$\Psi^{I}_2 (\Rho^{I}_5 \times \Rho^{I}_6)$
			&
			\begin{tikzpicture}[baseline={([yshift=-.5ex]current bounding box.center)}, node distance=1cm]
			\node (U) at (-.15, 0) {U};
			\node (P) at (0.5, 0) {P};
			\node (T) at (1.15, -0.3) {T};
			\node (L) at (1.15, 0.3) {L};
			\node (P2) at (1.8, 0) {P};
			\node (U2) at (2.45, 0) {U};
			
			\draw[->, to path={|- (\tikztotarget)}]
			(U) edge (P) (P) edge (L) (P) edge (T);
			\draw[<-, to path={-| (\tikztotarget)}]
			(L) edge (P2) (T) edge (P2);
			\draw[<-, to path={-| (\tikztotarget)}]
			(P2) edge (U2.west);
			
			\end{tikzpicture}&User $\xrightarrow{write}$  
			\begin{tikzpicture}[baseline={([yshift=-.5ex]current bounding box.center)}, node distance=1cm, auto,]
			\node[] (dummy) {};
			\coordinate (CENTER) at ($(dummy)$);
			\node[inner sep=0pt,above=0.0cm of dummy] (location) {Location};
			\node[inner sep=0pt,below=0.0cm of dummy] (time) {Timestamp};
			
			\node[right=of dummy] (post-right) {Post}
			edge[->, bend right=10] node[midway,above] {{\tiny $checkin$}} (location.east) 
			edge[->, bend left=10] node[midway,below] {{\tiny $at$}} (time.east); 
			
			\node[left=of dummy] (post-left) {Post}
			edge[->, bend left=10] node[midway,above] {{\tiny $checkin$}} (location.west)
			edge[->, bend right=10] node[midway,below] {{\tiny $at$}} (time.west);
			\end{tikzpicture}
			$\xleftarrow{write}$ User
			&\makecell[l]{Common Attributes}\\
			
			$\Psi^{I}_3 (\Rho^{I}_1 \times \Rho^{I}_5 \times \Rho^{I}_6)$
			&
			\begin{tikzpicture}[baseline={([yshift=-.5ex]current bounding box.center)}, node distance=1cm]
			\node (U11) at (-.15, 0.5) {};
			\node (U) at (-.15, 0) {U};
			\node (P) at (0.5, 0) {P};
			\node (W) at (1.15, -0.2) {T};
			\node (L) at (1.15, 0.2) {L};
			\node (P2) at (1.8, 0) {P};
			\node (U2) at (2.45, 0) {U};
			\node (U22) at (2.45, 0.5) {};
			
			\draw[->, to path={|- (\tikztotarget)}]
			(U) edge (P) (P) edge (L) (P) edge (W) (U) ;
			\draw[-, to path={|- (\tikztotarget)}]
			(U) edge (U11) ;
			\draw[<-, to path={|- (\tikztotarget)}]
			(U2) edge (U22.west);
			\draw[<-, to path={-| (\tikztotarget)}]
			(L) edge (P2) (W) edge (P2);
			\draw[<-, to path={-| (\tikztotarget)}]
			(P2) edge (U2.west);
			\draw[-, to path={-| (\tikztotarget)}]
			(U11) edge (U22.west);
			\draw[-, to path={|- (\tikztotarget)}]
			(U11) edge (U22.west);
			\end{tikzpicture}
			
			&
			\begin{tikzpicture}[baseline={([yshift=-.5ex]current bounding box.center)}, node distance=1cm, auto,]
			\node[] (dummy) {};
			\coordinate (CENTER) at ($(dummy)$);
			\node[inner sep=0pt,above=0.0cm of dummy] (location) {Location};
			\node[inner sep=0pt,below=0.0cm of dummy] (time) {Timestamp};
			
			\node[left=of dummy] (post-left) {Post}
			edge[->, bend left=10] node[midway,above] {{\tiny $checkin$}} (location.west)
			edge[->, bend right=10] node[midway,below] {{\tiny $at$}} (time.west);
			
			\node[right=of dummy] (post-right) {Post}
			edge[->, bend right=10] node[midway,above] {{\tiny $checkin$}} (location.east) 
			edge[->, bend left=10] node[midway,below] {{\tiny $at$}} (time.east); 
			
			
			\node[left=of post-left] (user-left) {User}
			edge[->] node[midway,above] {{\tiny $write$}} (post-left.west);
			
			\node[right=of post-right] (user-right) {User}
			edge[->] node[midway,above] {{\tiny $write$}} (post-right.east);
			
			\node[left=of post-left] (user-left) {User}
			edge[->, bend left=18] node[midway,above] {{\tiny $follow$}} (user-right);
			\end{tikzpicture}
			&\makecell[l]{Common Attributes \& Follower and Followee}\\

			\hline
			
		\end{tabular}
	}
	\vspace{-12pt}
\end{table*}
\subsection{Intra-Network Meta Diagram}\label{sec:intra}
The \textit{intra-network meta diagrams} can be defined in a similar way as \textit{inter-network meta diagrams} in Section~\ref{sec:inter}. There main differences lie in: \textit{inter-network meta diagrams} connect two nodes across two networks but \textit{intra-network meta diagrams} exist in one single network. Formally, we can define \textit{intra-network meta diagrams} as:

\noindent \textbf{Definition 5}
(Intra-Network Meta Diagram): Given attributed heterogeneous social network shcema $S_{{G}} = (\mathcal{N}_{\mathcal{V}} \cup \mathcal{N}_{\mathcal{T}},\mathcal{R}_{\mathcal{E}} \cup \mathcal{R}_{\mathcal{A}})$. An \textit{inter-network meta diagram} can be defined as a directed acyclic subgraph $\Psi^{I} = (\mathcal{N}_{\Psi}, \mathcal{R}_{\Psi}, N_s, N_t)$, where $\mathcal{N}_{\Psi} \subset (\mathcal{N}_{\mathcal{V}} \cup \mathcal{N}_{\mathcal{T}})$ and $\mathcal{R}_{\Psi} \subset (\mathcal{R}_{\mathcal{E}} \cup \mathcal{R}_{\mathcal{A}})$, while $N_s, N_t$ denote the source and target node types.

We only consider \textit{intra-network meta diagrams} which $N_s, N_t \in \{\mbox{U}\}$. We classify intra-network meta diagrams and represent the stacking process in the same way as inter-network meta diagrams. We list the notaion and physical meanings of \textit{intra-network meta paths} used in this paper in the first section of Table~\ref{tab:intra_meta_diagram}. Besides, several \textit{intra-network meta diagram} examples are presented in the second section of Table~\ref{tab:intra_meta_diagram}.
Similar to the \textit{inter-network meta diagrams}, \textit{intra-network meta diagrams} can be represented as $\Psi^{I} = \Rho^{I} \cup \Psi^{I}_{f^2} \cup \Psi^{I}_{a^2} \cup \Psi^{I}_{f,a} \cup \Psi^{I}_{f,a^2} \cup \Psi^{I}_{f^2,a^2}$.

Meta path \cite{SHYYW11} is a special type of the meta diagram in the shape of the path. In the following sections, we will directly use the term \textit{meta diagram} to refer to both \textit{meta path} and \textit{meta diagram}.
\section{Proposed Method}\label{sec:method}

{\our} is a general network alignment framework and the structure of {\our} is shown in Figure~\ref{fig:framework}. In {\our}. \textit{intra-network meta diagrams} will be applied to measure the proximity among users in single network and \textit{inter-network meta diagrams} will be utilized to calculate the proximity among user accounts from different networks. {\our} is a two-stage framework involving \textit{network synergistic partition} and \textit{parallel sub-network alignment}. \textit{Partitioned networks matching} acts as a bridge between these two stages. We will introduce three parts respectively in this section.
\subsection{Network Synergistic Partition}
The first stage of {\our} is \textit{network synergistic partition}, and we both exploit information within and across networks to 
obtain the optimal sub-networks. We measure the proximity among users within single network based on \textit{intra-network meta diagrams} and adjust sub-network structures synergistically with the support of \textit{inter-network meta diagrams}.

\subsubsection{Intra-Network Meta Diagram based Partition}\label{sec:intra-cluster}
%
We define IntraMD-Pro to measure the proximity among users in a heterogeneous social network.

\noindent \textbf{Definition 6}
(IntraMD-Pro): Given $\mathcal{D}_{\Psi_i^{I}}(x, y)$ to represent the set of diagram\ $\Psi^{I}_{i}$ starting from $x$ to $y$, and $\mathcal{D}_{\Psi_i^{I}}(x, \cdot)$ to represent the set of diagram\ $\Psi^{I}_{i}$ which go from $x$ to other nodes in the network. The IntraMD-Pro of node pair $(x, y)$ can be defined as
$$\mbox{IntraMD-Pro}(x, y) = \sum_i \omega_i \left(\frac{\left| \mathcal{D}_{\Psi_i^{I}}(x, y) \right| + \left| \mathcal{D}_{\Psi_i^{I}}(y, x) \right|} {{\left| \mathcal{D}_{\Psi_i^{I}}(x , \cdot) \right| } + \left| \mathcal{D}_{\Psi_i^{I}}(y , \cdot) \right|}\right),$$     
where $\omega_i$ is the weight of $\Psi^{I}_{i}$ and $\sum_i \omega_i = 1$. 

Accoriding to \cite{ZY15-2}, the specific values of hyperparameters $\omega_i$ can be adjusted automatically by optimizing certain learning objectives, e.g., clustering entropy as used in \cite{ZY15-2}. We will not elaborate the hyperparameter adjustment algorithm in this paper. What's more, we use $\mathbf{A}_i$ as the \textit{adjacency matrix} which represents $\Psi^{I}_{i}$ among users in the network. The proximity score matrix among users of $\Psi^{I}_{i}$ can be represented as
$\mb{S}_i = \mb{B}_i \circ \left( \mb{A}_i + \mb{A}_i^{\mathsf{T}} \right)$, where the matrix $\mb{B}_i$ represents the out-degree of user $x$ and $y$, e.g., $\mb{B}_i{(x,y)} = \left(\sum_m \mb{A}_i{(x,m)} + \sum_m \mb{A}_i{(y,m)}\right)^{-1}$. The $\circ$ symbol represents the Hadamard product. IntraMD-Pro matrix of the network can be represented as follows:
\begin{align*} 
\mathbf{S} = \sum_i \omega_i \mb{S}_i = \sum_i \omega_i \left( \mb{B}_i \circ \left( \mb{A}_i + \mb{A}_i^{\mathsf{T}} \right) \right). 
\end{align*}

We can represent the user-cluster belonging confidence scores as a vector $\mb{h}_i = (h_{i,1}, h_{i,2}, \ldots, h_{i,k})$, where $h_{i,j}$ denotes the confidence score that $u_i \in \mathcal{U}$ is in the sub-network $U_j \in \mathcal{C}$ ($\mathcal{C}$ is the set of detected clusters), and $k$ is the number of detected communities. Therefore, we can define the partition results of all users in $\mathcal{U}$ as the \textit{user-cluster belonging confidence matrix} $\mb{H}$, where $\bf{H} = [\bf{h}_1,$ $\bf{h}_2,$ $\ldots,$ $\bf{h}_n]^{\mathsf{T}}$ and $n = |\mathcal{U}|$. We choose to solve the following objective function to minimize the \textit{normalized-cut} ($Ncut$) cost \cite{SM00,Luxburg07} and achieve the optimal partition result:
\begin{align*}
\min_{\mb{H}}\ \ &\mbox{Tr} (\mb{H}^{\mathsf{T}}\mb{L}\mb{H}),\\
s.t. \ \ &\mb{H}^{\mathsf{T}}\mb{D}\mb{H} = \mb{I}.
\end{align*}
where the Laplacian matrix $\mb{L} = \mb{D} - \mb{S}$, the diagonal matrix $\mb{D}$ has $\mb{D}(i,i) = \sum_j \mb{S}(i,j)$ on its diagonal, and $\mb{I}$ is an identity matrix. 
\subsubsection{Inter-Network Meta Diagram based Partition}\label{sec:inter-cluster}
With the help of \textit{inter-network meta diagrams}, we can represent the extra knowledge about the aligned attributed heterogeneous networks from a more complete and convincing view. 

\textit{Inter-network meta diagrams} effectively indicate the closeness among the users across different networks, which can be quantified with the proximity scores in this paper. Given a pair of users $u_x^{(1)}$ and $u_y^{(2)}$, we denote the set of inter-network diagram $\Psi_i^{A}$ connecting $u_x^{(1)}$ and $u_y^{(2)}$ as $\mathcal{D}_{\Psi_i^{A}}(u_x^{(1)}, u_y^{(2)})$. Formally, we represent all inter-network meta diagram instances going out from user $u_x^{(1)}$ (or going into $u_y^{(2)}$) as set $\mathcal{D}_{\Psi_i^{A}}(u_x^{(1)}, \cdot)$ (or $\mathcal{D}_{\Psi_i^{A}}(\cdot, u_y^{(2)})$). The proximity score between $u_x^{(1)}$ and $u_y^{(2)}$ based on $\Psi_i^{A}$ can be defined as the following \textit{InterMD-Pro}.

\noindent \textbf{Definition 7}
(InterMD-Pro): Based on $\Psi_i^{A}$, the proximity between $u_x^{(1)}$ and $u_y^{(2)}$ in $\mathcal{G}$ can be represented as
$$s_{\Psi_i^{A}}(u_x^{(1)}, u_y^{(2)}) =\frac{2|\mathcal{D}_{\Psi_i^{A}}(u_x^{(1)}, u_y^{(2)})|}{|\mathcal{D}_{\Psi_i^{A}}(u_x^{(1)}, \cdot)| + |\mathcal{D}_{\Psi_i^{A}}(\cdot, u_y^{(2)})|}.$$
Based on the promixity of every single inter-network meta diagram, InterMD-proximity between users $u_x^{(1)}$ and 
$u_y^{(2)}$ in $\mathcal{G}$ can be defined as
$$\mbox{InterMD-Pro}(u_x^{(1)}, u_y^{(2)}) =  \sum_i \omega_i \left(s_{\Psi_i^{A}}(u_x^{(1)}, u_y^{(2)})\right).$$
where $\omega_i$ is the weight of \ $\Psi^{A}_{i}$ and $\sum_i \omega_i = 1$. 

\begin{figure}[t]
	\centering
	\begin{minipage}[l]{1.0\columnwidth}
		\centering
		\includegraphics[width=\textwidth]{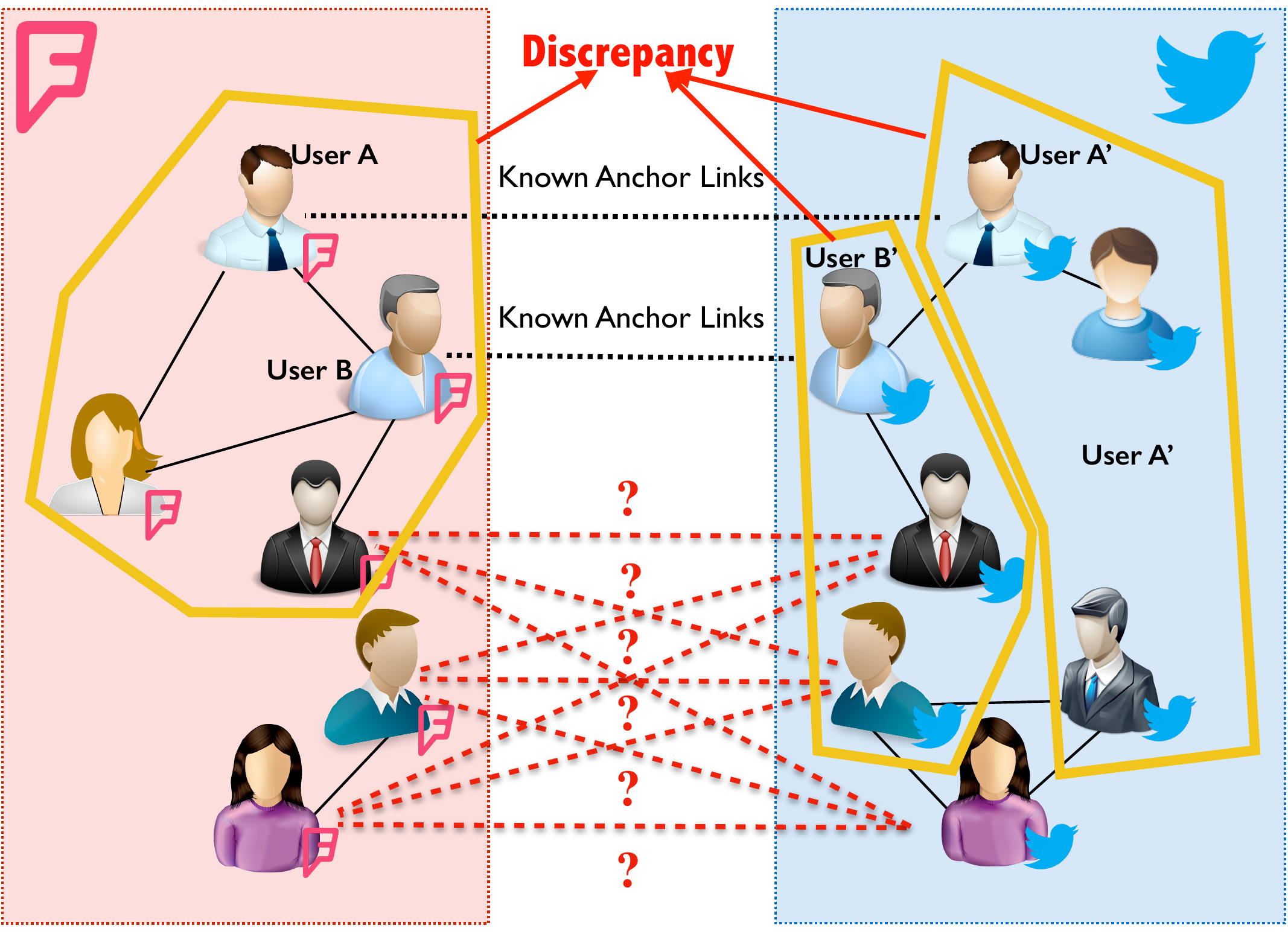}
	\end{minipage}
	\vspace{-5pt}
	\caption{An illustration of Discrepancy. There exists a known anchor link between User A and User A', also User B and User B'. If User A and User B are partitioned into the same sub-network, but User A' and User B' are partitioned into different sub-networks, the discrepancy will arise. Synergistic partition works on eliminating the discrepancy}\label{fig:discrepancy}\vspace{-10pt}
\end{figure}

The promixity matrix among all users across networks can be represented as $\mb{S}_{\Psi^{A}} \in \mathbb{R}^{|\mathcal{U}^{(1)}| \times |\mathcal{U}^{(2)}|}$, and $\mb{S}_{\Psi^{A}}(x,y) = \mbox{InterMD-Pro}(u_x^{(1)}, u_y^{(2)})$. 
We can correlate users together with their cluster belonging relationships effectively across networks with the matrix $\mb{S}_{\Psi^{A}}$. 
Given one user $u_l^{(1)}$ in $G^{(1)}$, 
we are able to calculate the user-cluster belonging confidence scores of $u_l^{(1)}$ with the user-cluster belonging confidence scores from $G^{(2)}$. Formally, we define \textit{Transition User-cluster Belonging Confidence Scores} as follow:
\begin{align*}
\bar{\mb{h}}_l^{(1)} = \sum_{j=1}^{|\mathcal{U}^{(2)}|} \mb{S}_{\Psi^{A}}(l,j) \cdot \mb{h}^{(2)}_j
\end{align*}
By maximizing the consensus of partition results based on the \textit{transition user-cluster belonging confidence scores}, we can refine the partition results with information from the other partially aligned network synergistically. In this paper, we will propose the definition of \textit{discrepancy}, which measures how different the shared user pairs are clustered across networks. We provide an illustration about \textit{discrepancy} in Figure~\ref{fig:discrepancy}. 

\noindent \textbf{Definition 8}
(Discrepancy): 	
Given two users $u_l^{(1)}$ and $u_m^{(1)}$ in $G^{(1)}$,  
If users $u^{(1)}_l$ and $u^{(1)}_m$ are partitioned into the same sub-network in $G^{(1)}$ but into different sub-networks based on the \textit{transition user-cluster belonging confidence scores} from $G^{(2)}$, then it will lead to the \textit{discrepancy} between the partition results of $u^{(1)}_l$, $u^{(1)}_m$. The confidence that $u^{(1)}_l$ and $u^{(1)}_m$ are in the same sub-network can be denoted as $\mb{h}_l^{(1)} (\mb{h}_m^{(1)})^{\top}$. Formally, the \textit{discrepancy} of  $u^{(1)}_l$ and $u^{(1)}_m$ is defined to be $d_{lm}(\mathcal{C}^{(1)}) = \left(\mb{h}_l^{(1)} (\mb{h}_m^{(1)})^{\top} - \bar{\mb{h}}^{(1)}_l (\bar{\mb{h}}^{(1)}_m)^{\top}\right )^2$
. Furthermore, the discrepancy of $\mathcal{C}^{(1)}$:
\begin{align*}
d(\mathcal{C}^{(1)}) &= \sum_i^{|\mathcal{U}^{(1)}|} \sum_{j=i+1}^{{|\mathcal{U}^{(1)}|}} d_{ij}(\mathcal{C}^{(1)})\\
\end{align*}
%

With the \textit{user-cluster belonging confidence matrices} $\mb{H}^{(1)}$ and $\mb{H}^{(2)}$,
the discrepancy of the aligned attributed heterogeneous networks $\mathcal{G}$ is
\begin{small}
	\begin{align*}
	\tiny
	d(\mathcal{C}^{(1)},\mathcal{C}^{(2)}) &= d(\mathcal{C}^{(1)})+d(\mathcal{C}^{(2)})\\
	&= \left \| \bar{\mb{H}}^{(1)} \left(\bar{\mb{H}}^{(1)} \right )^{\mathsf{T}} - \mb{H}^{(1)} \left ( \mb{H}^{(1)} \right)^{\mathsf{T}} \right \|^2_F \\
	&\ \ \ \ + \left \| \bar{\mb{H}}^{(2)} \left(\bar{\mb{H}}^{(2)} \right )^{\mathsf{T}} - \mb{H}^{(2)} \left (\mb{H}^{(2)} \right)^{\mathsf{T}} \right \|^2_F.
	\end{align*} 
\end{small}
Where $\bar{\mb{H}}^{(1)} = [\bar{\bf{h}}^{(1)}_1,$ $\bar{\bf{h}}^{(1)}_2,$ $\ldots,$ $\bar{\bf{h}}^{(1)}_n]^{\top}$ and $n = |\mathcal{U}^{(1)}|$, and $\bar{\mb{H}}^{(2)}$ is the same situation. Besides, $\bar{\mb{H}}^{(2)}=\left(\mb{S}_{\Psi^{A}}\right)^{\top}\mb{H}^{(1)}$ and $\bar{\mb{H}}^{(1)}=\mb{S}_{\Psi^{A}}\mb{H}^{(2)}$.
\subsubsection{Synergistic Partition of Multiple Networks}\label{sec:joint-cluster}
By taking both \textit{Intra-Network Meta Diagram based Partition} and \textit{Inter-Network Meta Diagram based Partition} into considerations, the optimal synergistic partition results $\mathcal{{C}}^{(1)}$ and $\mathcal{{C}}^{(2)}$
can be achieved by minimizing both the Ncut costs and the discrepancy simultaneously as follows:
\begin{align*}
\arg \min_{\mathcal{C}^{(1)}, \mathcal{C}^{(2)}} \alpha \cdot Ncut(\mathcal{C}^{(1)}) + \beta \cdot Ncut(\mathcal{C}^{(2)}) + \theta \cdot d(\mathcal{C}^{(1)}, \mathcal{C}^{(2)})
\end{align*}
where $\alpha$, $\beta$ and $\theta$ represent the weights of these compositions. We can replace $Ncut(\mathcal{C}^{(1)})$, $Ncut(\mathcal{C}^{(2)})$, $d(\mathcal{C}^{(1)}, \mathcal{C}^{(2)})$ with the terms derived before, and the joint objective function can be rewrited as:
\begin{small}
	\begin{align*}
	\min_{\mb{H}^{(1)}, \mb{H}^{(2)}}\ \ &\alpha \cdot \mbox{Tr} (({\mb{H}^{(1)}})^{\mathsf{T}}\mb{L}^{(1)}\mb{H}^{(1)}) + \beta \cdot \mbox{Tr} (({\mb{H}^{(2)}})^{\mathsf{T}}\mb{L}^{(2)}\mb{H}^{(2)}) \\
	&+ \theta \cdot \left \| \bar{\mb{H}}^{(1)} \left(\bar{\mb{H}}^{(1)} \right )^{\mathsf{T}} - \mb{H}^{(1)} \left ( \mb{H}^{(1)} \right)^{\mathsf{T}} \right \|^2_F\\
	&+ \theta \cdot \left \| \bar{\mb{H}}^{(2)} \left(\bar{\mb{H}}^{(2)} \right )^{\mathsf{T}} - \mb{H}^{(2)} \left (\mb{H}^{(2)} \right)^{\mathsf{T}} \right \|^2_F,\\
	&s.t. \ \ ({\mb{H}^{(1)}})^{\mathsf{T}}\mb{D}^{(1)}\mb{H}^{(1)} = \mb{I}^{(1)}, ({\mb{H}^{(2)}})^{\mathsf{T}}\mb{D}^{(2)}\mb{H}^{(2)} = \mb{I}^{(2)},
	\end{align*}
\end{small}
The joint objective function involves two variables: $\mb{H}^{(1)}$ and $\mb{H}^{(2)}$, and the objective is not jointly convex. Besides, the objective function contains complex orthogonality constraints which are numerically expensive to preserve in optimization. In order to preserve constraints in an efficient way during the learning process, we propose to relax the objective function as follows:
\begin{small}
	\begin{align*}
	&\min_{\mb{H}^{(1)}, \mb{H}^{(2)}}\ \ \alpha \cdot Ncut(\mathcal{C}^{(1)}) + \beta \cdot Ncut(\mathcal{C}^{(2)}) + \theta \cdot d(\mathcal{C}^{(1)}, \mathcal{C}^{(2)})\\
	&+ \rho_{1} \left \|({\mb{H}^{(1)}})^{\mathsf{T}}\mb{D}^{(1)}\mb{H}^{(1)}-\mb{I}^{(1)} \right \|^2_F +\rho_{2} \left \|({\mb{H}^{(2)}})^{\mathsf{T}}\mb{D}^{(2)}\mb{H}^{(2)}-\mb{I}^{(2)} \right \|^2_F,
	\end{align*}
\end{small}
By setting $\rho_1$ and $\rho_2$ with large values, e.g., $10^9$, optimizing the above function is (approximately) equivalent to the original objective function. We design an hierarchical alternative variable updating process for solving the problem:

\noindent $\bullet$  \textbf{Step (1)}: Fix $\mb{H}^{(2)}$, Update $\mb{H}^{(1)}$.

With $\mb{H}^{(2)}$ fixed, the objective function involving $\mb{H}^{(1)}$ is:
\begin{small}
	\begin{align*}
	\mathcal{L}= \min_{\mb{H}^{(1)}}\ \ &\alpha \cdot \mbox{Tr} (({\mb{H}^{(1)}})^{\mathsf{T}}\mb{L}^{(1)}\mb{H}^{(1)})+ \theta \cdot \left \| \mb{X} - \mb{H}^{(1)} \left ( \mb{H}^{(1)} \right)^{\mathsf{T}} \right \|^2_F \\
	&+ \theta \cdot \left \| \mb{S}^{\mathsf{T}}_{\Psi^{A}}\mb{H}^{(1)}\left ( \mb{H}^{(1)} \right)^{\mathsf{T}}\mb{S}_{\Psi^{A}}  - \mb{Y} \right \|^2_F \\
	&+ \rho_{1} \left \|({\mb{H}^{(1)}})^{\mathsf{T}}\mb{D}^{(1)}\mb{H}^{(1)}-\mb{I}^{(1)} \right \|^2_F,
	\end{align*}
\end{small}
where $\mb{X} = \mb{S}_{\Psi^{A}}\mb{Y}\mb{S}^{\mathsf{T}}_{\Psi^{A}}$ and $\mb{Y} = \mb{H}^{(2)}\left ( \mb{H}^{(2)} \right)^{\mathsf{T}}$. Based on the Gradient Descent, we calculate 
$\mb{H}^{(1)}_k$ representing $\mb{H}^{(1)}$ after $k$ descent steps:
\begin{align*}
\mb{H}^{(1)}_k = \mb{H}^{(1)}_{k-1} - \eta_1 \cdot \nabla\mathcal{L}(\mb{H}^{(1)}) 
\end{align*}
$\eta_1$ is the step length, and the gradient $\nabla\mathcal{L}(\mb{H}^{(1)})$ is:
\begin{small}
	\begin{align*}
	\nabla&\mathcal{L}(\mb{H}^{(1)}) = \frac{\partial\mathcal{L}}{\partial\mb{H}^{(1)}} \\&+\alpha \cdot \left(\mb{L}^{(1)}\mb{H}^{(1)} + (\mb{L}^{(1)})^{\mathsf{T}}\mb{H}^{(1)}\right) \\&+ 2\theta\left(\mb{X}\mb{H}^{(1)}+\mb{X}^{\mathsf{T}}\mb{H}^{(1)}+2\mb{H}^{(1)}(\mb{H}^{(1)})^{\mathsf{T}}\mb{H}^{(1)} \right)\\&+
	2\theta\left(2\mb{S}_{\Psi^{A}}\mb{S}^{\mathsf{T}}_{\Psi^{A}}\mb{H}^{(1)}(\mb{H}^{(1)})^\mathsf{T}\mb{S}_{\Psi^{A}}\mb{S}^{\mathsf{T}}_{\Psi^{A}}\mb{H}^{(1)} - \mb{X}\mb{H}^{(1)}-\mb{X}^{\mathsf{T}}\mb{H}^{(1)} \right)\\&+
	4\rho_{1}\left(\mb{D}^{(1)}\mb{H}^{(1)}(\mb{H}^{(1)})^{\mathsf{T}}\mb{D}^{(1)}\mb{H}^{(1)}-\mb{D}^{(1)}\mb{H}^{(1)}\right)
	\end{align*}
\end{small}
\noindent $\bullet$  \textbf{Step (2)}: Fix $\mb{H}^{(1)}$, Update $\mb{H}^{(2)}$.

When $\mb{H}^{(1)}$ is fixed, we have the objective function $\mathcal{L}$ as follows:
\begin{small}
	\begin{align*}
	\mathcal{L} = &\min_{\mb{H}^{(2)}}\ \ \beta \cdot \mbox{Tr} (({\mb{H}^{(2)}})^{\mathsf{T}}\mb{L}^{(2)}\mb{H}^{(2)})+ \theta \cdot \left \| \mb{S}_{\Psi^{A}}\mb{H}^{(2)}\left ( \mb{H}^{(2)} \right)^{\mathsf{T}}\mb{S}^{\mathsf{T}}_{\Psi^{A}}  - \mb{Y} \right \|^2_F \\
	&+ \theta \cdot \left \| \mb{X} -\mb{H}^{(2)}\left ( \mb{H}^{(2)} \right)^{\mathsf{T}} \right \|^2_F + \rho_{2} \left \|({\mb{H}^{(2)}})^{\mathsf{T}}\mb{D}^{(2)}\mb{H}^{(2)}-\mb{I}^{(2)} \right \|^2_F,
	\end{align*}
\end{small}
where $\mb{X} = \mb{S}^{\mathsf{T}}_{\Psi^{A}}\mb{Y}\mb{S}_{\Psi^{A}}$ and $\mb{Y} = \mb{H}^{(1)}\left ( \mb{H}^{(1)} \right)^{\mathsf{T}}$. 
The method of updating $\mb{H}^{(2)}$ with a fixed $\mb{H}^{(1)}$ is almost the same as \textbf{Step (1)}:
\begin{small}
	\begin{align*}
	\mb{H}^{(2)}_k = \mb{H}^{(2)}_{k-1} - \eta_2 \cdot \nabla\mathcal{L}(\mb{H}^{(2)}) 
	\end{align*}
\end{small}
We will iteratively operate \textbf{Step (1)} and \textbf{Step (2)}, and every iteration will operate one step descent for $\mb{H}^{(1)},\mb{H}^{(2)}$ until convergence.

Based on the learned matrix $\mb{H}^{(1)}$, we can learn the clusters of users in network $G^{(1)}$ by applying K-Means algorithms to the learned latent vectors, i.e., rows of matrix $\mb{H}^{(1)}$, and the detected clusters can be represented as set $\mathcal{C}^{(1)}=\{U^{(1)}_1, U^{(1)}_2, \ldots, U^{(1)}_k\}$. For the users within the same cluster, we propose to extract a sub-network formed by these users and other associated nodes/attributes. For instance, based on the cluster $U^{(1)}_l \in \mathcal{C}^{(1)}$, we can represent the extracted sub-network as $G^{(1)}_l$. Formally, the set of extracted sub-network from $G^{(1)}$ based on the clustering result $\mathcal{C}^{(1)}$ can be represented as $\mathcal{G}^{(1)}=\{G^{(1)}_1,G^{(1)}_2,\dots,G^{(1)}_k\}$. 
Here, we need to add a remark that the synergistic network partition process involves an iterative variable updating process, which may take some time to converge. Meanwhile, in the real-world application of the proposed model, such a step can be done in an offline manner, where the clustering results can be computed and stored in hard-drive in advance. It will greatly improve the learning efficiency of {\our} in aligning the large-scale social networks. 

\subsection{Partitioned Networks Matching}\label{sec:subnetwork_matching}
After partitioning the original networks, it's critical to matching the sub-networks from different networks, which is the prerequisite for the next stage. As a bridge, the matching step should consider not only the object of \textit{network synergistic partition} but also the target of \textit{parallel sub-network alignment}. Here, we propose the sub-network Matching Score as the metric to serve for partitioned networks matching in {\our}.
\noindent \textbf{Definition 9}
(Matching Score): Given two sub-network $G^{(1)}_i$ and $G^{(2)}_j$, which comes from $\mathcal{G}^{(1)}$ and $\mathcal{G}^{(2)}$ respectively. We define Matching Score(M-Score) between $G^{(1)}_i$ and $G^{(2)}_j$ as:
$$\mbox{M-Score}(G^{(1)}_i, G^{(2)}_j) = | \mathcal{A}(G^{(1)}_i, G^{(2)}_j|) \cdot \frac{| \mathcal{A}(G^{(1)}_i, G^{(2)}_j)|}{|U^{(1)}_i| \cdot |U^{(2)}_j|}$$
where $\mathcal{A}(G^{(1)}_i, G^{(2)}_j)$ is the set of known anchor links between $G^{(1)}_i$ and $G^{(2)}_j$, and $U^{(1)}_i$, $U^{(2)}_j$ are sets of user accounts belongs to $G^{(1)}_i$, $G^{(2)}_j$. 
In fact, the second term above is the proportion of known links of all links across $G^{(1)}_i$ and $G^{(2)}_j$. M-Score takes both the number of known anchor users and the performance of pruning negative links into considerations. We can match the sub-networks according to the descending rank of M-Score to achieve the sub-network matching results $\mathcal{M} = \{\mathcal{M}_1,\mathcal{M}_2, \dots,\mathcal{M}_s\}$, and $\mathcal{M}_i = \{G^{(1)}_a, G^{(2)}_b\}$. Here, $s$ is a parameter we set corresponding to the $top\ s (s \le k)$ pairs for alignment. Then {\our} will start to focus on \textit{parallel sub-network alignment} on $\mathcal{M}$.
\subsection{Sub-network Alignment}\label{sec:alignment}
In this part, we will introduce the alignment model for all the sub-network pairs in $\mathcal{M}$. In the following section, we will take $\mathcal{M}_i = \{G^{(1)}_a, G^{(2)}_b\}$ as an example to illustrate the alignment method and the alignment process on the remaining sub-network pairs is identical to the method introduced here. 
\subsubsection{Optimization Objective Function}
For all the potential anchor links between $G^{(1)}_a$ and $G^{(2)}_b$ in set $\mathcal{H}$ involving both the labeled and unlabeled anchor link instances, a set of features will be extracted based on \textit{inter-network meta diagrams}. Formally, the feature vector extracted for the link $l \in \mathcal{H}$ can be represented as vector $\mb{x}_l \in \mathbb{R}^{f}$, where $f$ is the number of types of inter-network meta diagrams. Meanwhile, we can denote the label of link $l \in \mathcal{H}$ as $y_l \in \mathcal{Y}$, and $\mathcal{Y} = \{0, +1\}$, which denotes the existence of anchor link $l$ between the networks. For the existing anchor links in set $\mathcal{D}$, they will be assigned with $+1$ label; while the labels of anchor links in $\mathcal{P}$ are unknown. All the labeled anchor links in set $\mathcal{D}$ can be represented as a tuple set $\{(\mb{x}_l, y_l)\}_{l \in \mathcal{D}}$. The \textit{discriminative} component can effectively differentiate the positive instances from the non-existing ones, which can be denoted as mapping $f(\cdot; \mb{\theta}_f): \mathbb{R}^d \to \{+1, 0\}$ parameterized by $\mb{\theta}_f$. In this paper, we will use a linear model to fit the link instances, and the \textit{discriminative} model to be learned can be represented as $f(\mb{x}_l; {\mb{w}}) = \mb{w}^\top \mb{x}_l + b$, where $\mb{\theta}_f = [\mb{w}, b]$. By adding a dummy feature $1$ for all the anchor link feature vectors, we can incorporate bias term $b$ into the weight vector $\mb{w}$ and the parameter vector can be denoted as $\mb{\theta}_f = \mb{w}$ for simplicity. The introduced \textit{discriminative} loss function on the labeled set $\mathcal{D}$ can be represented as
\begin{align*}
L(f, \mathcal{D} ; \mb{w}) = \sum_{l \in \mathcal{D} } \big( f(\mb{x}_l; \mb{w}) - y_l \big)^2 = \sum_{l \in \mathcal{D} } (\mb{w}^\top \mb{x}_l - y_l)^2.
\end{align*}
Meanwhile, we also propose to utilize the unlabeled anchor links to encourage the learned model can capture the salient structures of all the anchor link instances. Based on the above discriminative model function $f(\cdot; \mb{w})$, for an unlabeled anchor link $l \in \mathcal{P}$, we can represent its inferred ``label'' as $y_l = f(\mb{x}_l; \mb{w})$. Considering that the result of $f(\cdot; \mb{w})$ may not necessary the exact label values in $\mathcal{Y}$, in the \textit{generative} component, we can represent the generated anchor link label as $sign\big(f(\mb{x}_l; \mb{w})\big) \in \{+1, 0\}$. How to determine its value will be introduced later in the joint function. The loss function introduced in the generative component based on the unlabeled anchor links can be denoted as 
\begin{align*}
L(f, \mathcal{P}; \mb{w}) = \sum_{l \in \mathcal{P}} \Big( \mb{w}^\top \mb{x}_l - sign\big(f(\mb{x}_l; \mb{w}) \big) \Big)^2.
\end{align*}
As introduced before, the anchor links to be inferred between networks are subject to the \textit{one-to-one} cardinality constraint. Subject to the cardinality constraint, the prediction tasks of anchor links between networks are no longer independent. For instance, if anchor link $(u_i^{(1)}, u_j^{(2)})$ is predicted to be positive, then all the remaining anchor links incident to $u_i^{(1)}$ and $u_j^{(2)}$ in the unlabeled set $\mathcal{P}$ will be negative by default. Viewed in such a perspective, the cardinality constraint on anchor links should be effectively incorporated in model building, which will be modeled as the mathematical constraints on node degrees. To represent the user node-anchor link relationships in networks $G_a^{(1)}$ and $G_b^{(2)}$ respectively, we introduce the user node-anchor link incidence matrices $\mb{A}^{(1)}  \in \{0, 1\}^{|\mathcal{U}_a^{(1)}| \times |\mathcal{H}|}, \mb{A}^{(2)} \in \{0, 1\}^{|\mathcal{U}_b^{(2)}| \times |\mathcal{H}|}$. Entry $A^{(1)}(i, j) = 1$ iff anchor link $l_j \in \mathcal{H}$ is connected with $u_i^{(1)}$ in $G_a^{(1)}$, and it is similar for $\mb{A}^{(2)}$.

According to the analysis provided before, we can represent the labels of links in $\mathcal{H}$ as vector $\mb{y} \in \{+1, 0\}^{|\mathcal{H}|}$, where entry $y(i)$ represents the label of link $l_i \in \mathcal{H}$. 
Based on the anchor link label vector $\mb{y}$, user node-anchor link incidence matrices $\mb{A}^{(1)}$ and $\mb{A}^{(2)}$, 
the \textit{one-to-one} constraint on anchor links can be denoted as the constraints on node degrees as follows:
\begin{alignat*}{2}
\mb{0} \le \mb{A}^{(1)} \mb{y} \le \mb{1}
\mbox{, and }
\mb{0} \le \mb{A}^{(2)} \mb{y} \le \mb{1}.
\end{alignat*}

By combining the loss terms introduced by the labeled and unlabeled anchor links together with the cardinality constraint, we can represent the joint optimization objective function as
\begin{align*}
\min_{\mb{w}, \mb{y}}\  &L(f, \mathcal{D} ; \mb{w}) + L(f, \mathcal{P}; \mb{w}) + \gamma \cdot \left\|\mb{w}\right\|_2^2\\
&s.t.\ \ \ \ \ \ y_l \in \{+1, 0\}, \forall l \in \mathcal{H},\\
&\ \ \ \ \ \  \mb{0} \le \mb{A}^{(1)} \mb{y} \le \mb{1} \mbox{, and } \mb{0} \le \mb{A}^{(2)} \mb{y} \le \mb{1}.
\end{align*}
In fact, we can simplify the loss function as:
\begin{align*}
&L(f, \mathcal{D} ; \mb{w}) + L(f, \mathcal{P}; \mb{w}) = L(f, \mathcal{H}; \mb{w}) =  \left\| \mb{X} \mb{w} - \mb{y} \right\|_2^2,
\end{align*}
where matrix $\mb{X} = [\mb{x}_{l_1}^\top, \mb{x}_{l_2}^\top, \cdots, \mb{x}_{l_{|\mathcal{H}|}}^\top]^T$ denotes the feature matrix of all the links in the set $\mathcal{H}$.

The objective function involves two variables, i.e., variable $\mb{w}$, label $\mb{y}$, and the objective is not jointly convex with regarding these variables. So obtaining their optimal solution will be NP-hard. In this paper, we design a hierarchical alternative variable updating process for solving the problem instead:

\noindent $\bullet$  \textbf{Step (1)}: Fix $\mb{y}$, Update $\mb{w}$.

With $\mb{y}$ fixed, the objective function involving $\mb{w}$ is:
\begin{align*}
\min_{\mb{w}} \frac{c}{2} \left\| \mb{X} \mb{w} - \mb{y} \right\|_2^2 + \frac{1}{2} \left\|\mb{w}\right\|_2^2.
\end{align*}

Here, the objective function is a quadratic convex function, and its optimal solution can be represented as
$$ \mb{w} = c(\mb{I} + c\mb{X}^\top \mb{X})^{-1} \mb{X}^\top \mb{y},$$
where $c(\mb{I} + c\mb{X}^\top \mb{X})^{-1} \mb{X}^\top$ is a constant matrix. Therefore, the weight vector $\mb{w}$ depends only on the $\mb{y}$ variable.

\noindent $\bullet$  \textbf{Step (2)}: Fix $\mb{w}$, Update $\mb{y}$.

With $\mb{w}$ fixed, together with the constraint, we know that terms $L(f, \mathcal{D} ; \mb{w})$, $L(f, \mathcal{P}; \mb{w})$ and $\left\| \mb{w} \right\|_2^2$ are all constant. And the objective function will be
\begin{align*}
&\min_{\mb{y}} \left\| \mb{X}\mb{w} - \mb{y} \right\|_2^2 \\
&s.t.\ \ y_l \in \{+1, 0\}, \forall l \in \mathcal{H},\\
&\ \ \ \ \ \  \mb{0} \le \mb{A}^{(1)} \mb{y} \le \mb{1} \mbox{, and } \mb{0} \le \mb{A}^{(2)} \mb{y} \le \mb{1}.
\end{align*}
It is an integer programming problem, which has been shown to be NP-hard and no efficiently algorithm exists that lead to the optimal solution. In this paper, we will use the greedy link selection algorithm proposed in \cite{ZCZCY17} based on values $\hat{\mb{y}} =  \mb{X}\mb{w}$, which has been proven to achieve $\frac{1}{2}$-approximation of the optimal solution.

\subsubsection{Parallel Implementation of Sub-network Alignment}

Sub-network alignment involves two iterative steps. The time complexity of these two steps is related to $|\mathcal{H}|$ which is determined by the number of users from two sub-networks. The alignment for all the sub-network pairs in the set  $\mathcal{M}$ can be implemented in parallel, so compared with the alignment method conducted in the whole networks directly, {\our} has the apparent advantage even counting the time consumption of the \textit{network synergistic partition}.
Finally, we have to aggregate alignment results from parallel sub-network alignment in the sub-network pairs. In {\our}, we choose to preserve the original results from all sub-network pairs in the set $\mathcal{M}$ as the final result.  
\section{Experiments}\label{sec:experiment}
To demonstrate the effectiveness of {\our}, extensive experiments have been done on real-world heterogeneous social networks. In this section, we describe the dataset first. Next, the experimental settings are introduced. Then we show the experimental results together with the convergence and time analysis. At last, we provide parameter sensitivity analysis.  

\subsection{Dataset}
Our dataset comes from two real-world heterogeneous networks: \textit{Foursquare} and \textit{Twitter}, which are both famous online social networks. The key statistical data of these two networks is listed in Table~\ref{tab:datastat}. Detailed information about the strategy of crawling the dataset can be reached in \cite{KZY13}.
\begin{table}[t]
	\caption{Properties of the Heterogeneous Networks}
	\label{tab:datastat}
	\centering
	\begin{tabular}{clrr}
		\toprule
		&&\multicolumn{2}{c}{network}\\
		\cmidrule{3-4}
		&property &\textbf{Twitter} &\textbf{Foursquare}   \\
		\midrule 
		\multirow{3}{*}{\# node}
		&user   & 5,223 & 5,392 \\
		&tweet/tip  & 9,490,707 & 48,756 \\
		&location & 297,182 & 38,921 \\
		\midrule 
		\multirow{2}{*}{\# link}
		&friend/follow    &164,920  &76,972 \\
		&write    & 9,490,707 & 48,756 \\
		\bottomrule
	\end{tabular}
	\vspace{-10pt}
\end{table}

\subsection{Experimental Settings}
\subsubsection{Experimental Setup}

\begin{table}[t]
	\caption{Performance comparison of different methods for Network Alignment.}
	\label{tab:main_result}
	\centering
	\setlength{\tabcolsep}{3pt}
	{
		\begin{tabular}{lrcccc}
			\toprule
			\multicolumn{2}{l}{}&\multicolumn{4}{c}{Metrics}\\
			\cmidrule{3-6}
			&Methods	&Precision	& Recall	& F1	 &Time (sec)\\
			\midrule
			&{\our}($\theta=10$)	&0.677$\pm$0.002     &0.500$\pm$0.001     &0.575$\pm$0.001     &\textbf{7.62}	\\
			&{\our}($\theta=80$)	&\textbf{0.691}$\pm$\textbf{0.001}     &\textbf{0.532}$\pm$\textbf{0.011}     &\textbf{0.601}$\pm$\textbf{0.006}     &9.12  	\\
			&{\our}($\theta=100$)	&0.684$\pm$0.010     &0.515$\pm$0.018    &0.588$\pm$0.015     &12.87  	\\
			\cmidrule{2-6}
			&{\sppu}&0.481$\pm$0.002       &0.392$\pm$0.013       &0.432$\pm$0.006        &450.14  \\
			&{\kmeanspu}	&0.415$\pm$0.003      &0.239$\pm$0.009       &0.303$\pm$0.008    &361.97   \\
			\cmidrule{2-6}
			&{\pusvm}	&0.318$\pm$0.004       &0.281$\pm$0.002       &0.298$\pm$0.003      &49393.81  	\\
			&{\svm}	&0.137$\pm$0.008       &0.259$\pm$0.003       &0.178$\pm$0.002         &6480.38  	\\
			&{\dw}	&0.043$\pm$0.001       &0.075$\pm$0.001       &0.054$\pm$0.000          &18756.13  	\\
			&{\mv}	&0.071$\pm$0.001       &0.102$\pm$0.002       &0.084$\pm$0.001          &21314.67  	\\

			\bottomrule
		\end{tabular}
	}
	
\end{table}
In the experiments, we can obtain the set of anchor links across Foursquare and Twitter, which will be the positive links. The links between users from Foursquare and Twitter except for anchor links can be treated as negative links. We apply the 2-fold cross-validation to partition the links with the ratio 1 : 1. One fold will be used as the training set and the other one will be treated as the test set. The features depending on the known anchor links like inter-network meta diagrams are extracted only on the basis of the training set.
All codes are implemented in Python 3, and we run the experiments on a Dell PowerEdge T630 Server with 2 20-core Intel CPUs and 256GB memory. The operating system is Ubuntu 16.04.3.
\subsubsection{Comparison Methods}
Comparison methods in the experiments can be divided into 2 categories according to whether original networks are partitioned or not in building models.

\noindent \textbf{Comparison Methods without partition}:
\begin{itemize}
	\item \textbf{\pusvm}: {\pusvm} extends the cardinality constrained link prediction model in \cite{ZCZCY17} by incorporating inter-network meta diagrams.
	\item \textbf{\svm}:  {\svm} extends the anchor links prediction model in \cite{KZY13} by incorporating inter-network meta diagrams as features.
	\item \textbf{\dw}: A random walk based network embedding method \cite{PAS14}, but it is designed to deal with the homogeneous network. We utilize it to learn the representation of users merely based on the friendship information and concatenate the representations of two users as the feature of a potential anchor link. Then SVM will be trained to predict anchor links based on this feature. 
	\item \textbf{\mv}: A meta-path based heterogeneous network embedding method \cite{DCS17}, but it can only handle specific one meta-path. Similar to {\dw}, we use it to learn the embedding of users and predict anchor links with a SVM. We report the best result of different intra-network meta diagrams. 
\end{itemize}
\noindent \textbf{Comparison Methods with partition}:
\begin{itemize}
	\item \textbf{\our}: {\our} is the model proposed in this paper.
	\item \textbf{\sppu}: It implements the network partition using spectral clustering, and the sub-network alignment algorithm is the same as {\our}.
	\item \textbf{\kmeanspu}: In {\kmeanspu}, we directly use k-means clustering to partition the networks.	
\end{itemize}

Some recent methods based on graph embedding and structural seeds like \cite{wang2018deepmatching,zhang2018mego2vec,kazemi2015growing,du2019joint} are designed for homogeneous graph and attributed networks, which are different from our problem definition. Therefore, we do not include all of them in comparison methods.

\subsection{Experimental Results with Analysis}
We will evaluate \textit{network synergistic partition} and \textit{parallel sub-network alignment} respectively together with \textit{partitioned networks matching} that connects them.

\subsubsection{Network Synergistic Partition}
\begin{figure}[t]

	\centering
	\begin{minipage}[l]{0.9\columnwidth}
		\centering
		\includegraphics[width=\textwidth]{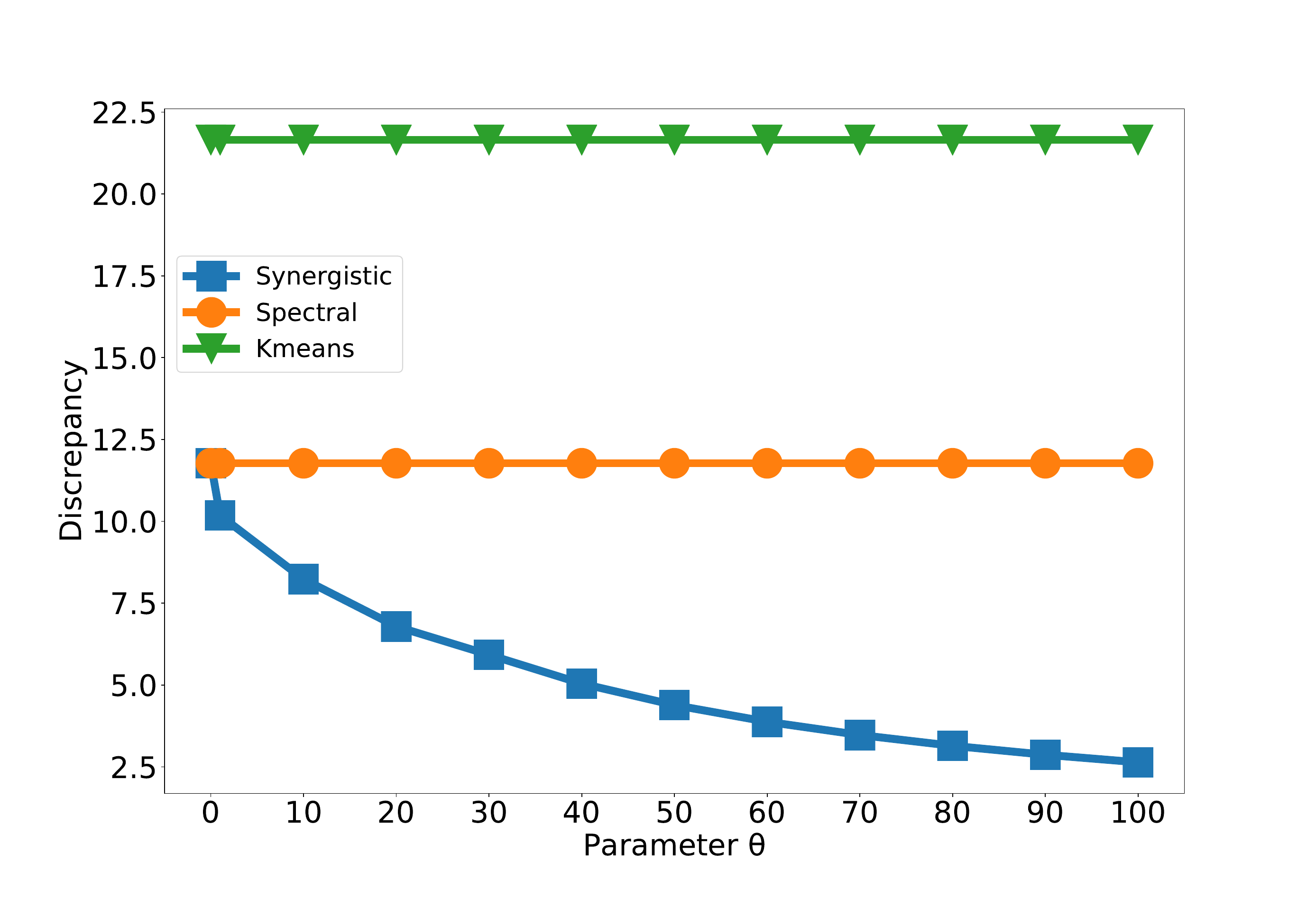}
	\end{minipage}
	\vspace{-10pt}
	\caption{Discrepancy with different parameter $\theta$}\label{fig:discrepancy_theta} 
\end{figure}
\begin{figure}[t]

	\centering
	\begin{minipage}[l]{0.9\columnwidth}
		\centering
		\includegraphics[width=\textwidth]{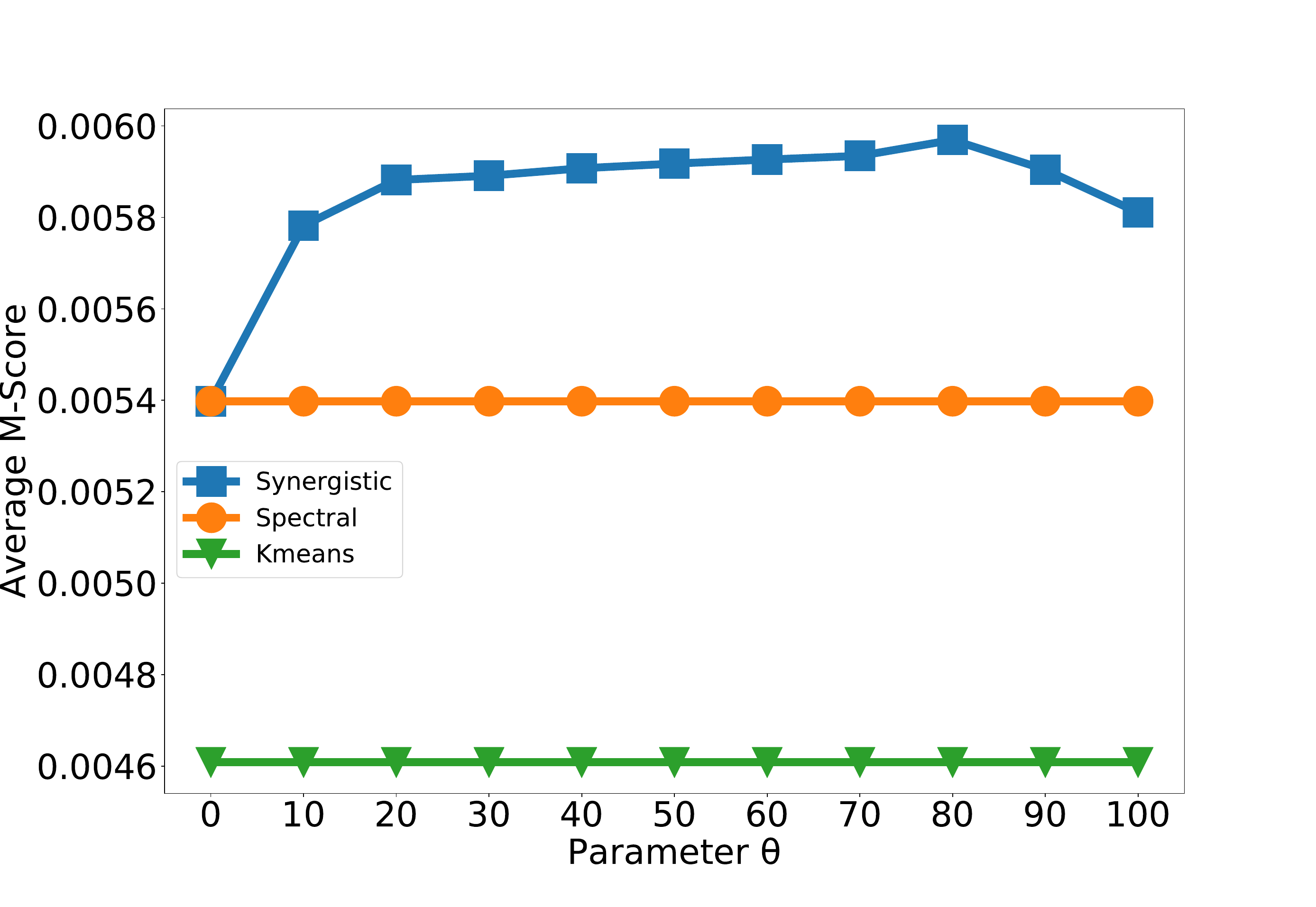}
	\end{minipage}
	\vspace{-10pt}
	\caption{The average M-Score of Top 20 sub-network pairs}\label{fig:avg_closeness_theta} 
\end{figure}
To illustrate the effectiveness of \textit{network synergistic partition}, we evaluate the results of the partition before performing the alignment. First, according to Definition 8, the discrepancy of different partition methods is displayed in Figure~\ref{fig:discrepancy_theta} which shows that \textit{network synergistic partition} can minimize the discrepancy compared to the other two methods, and the effect of decreasing the discrepancy becomes more apparent as the weight of $\theta$ increases. 

Besides, according to M-Score defined in Section~\ref{sec:subnetwork_matching}, we can observe the average M-Score of Top 20 subnetwork pairs in Figure~\ref{fig:avg_closeness_theta}.
It essentially demonstrates that \textit{network synergistic partition} has the best performance in the task of partitioning according to our requirements. Here we do not apply some classic metrics which are often used to evaluate the clustering result, because in {\our}, the partition is used by the next stage in order to better perform the alignment. Conventional metrics for clustering may not be effective here. For example, the partition method obtaining a better result based on conventional clustering metrics in every single network does not guarantee that partitioning multiple networks simultaneously can obtain well-matched sub-networks. From Figure~\ref{fig:discrepancy_theta} and Figure~\ref{fig:avg_closeness_theta}, we can find that as $\theta$ rises, the discrepancy is declining, but it does not bring the monotonous rise of the average M-Score. We will make detailed analysis through the discussion on the parameter $\theta$ in Section~\ref{sec:parameter}.

\begin{figure}[t]

	\centering
	\begin{minipage}[l]{0.9\columnwidth}
		\centering
		\includegraphics[width=\textwidth]{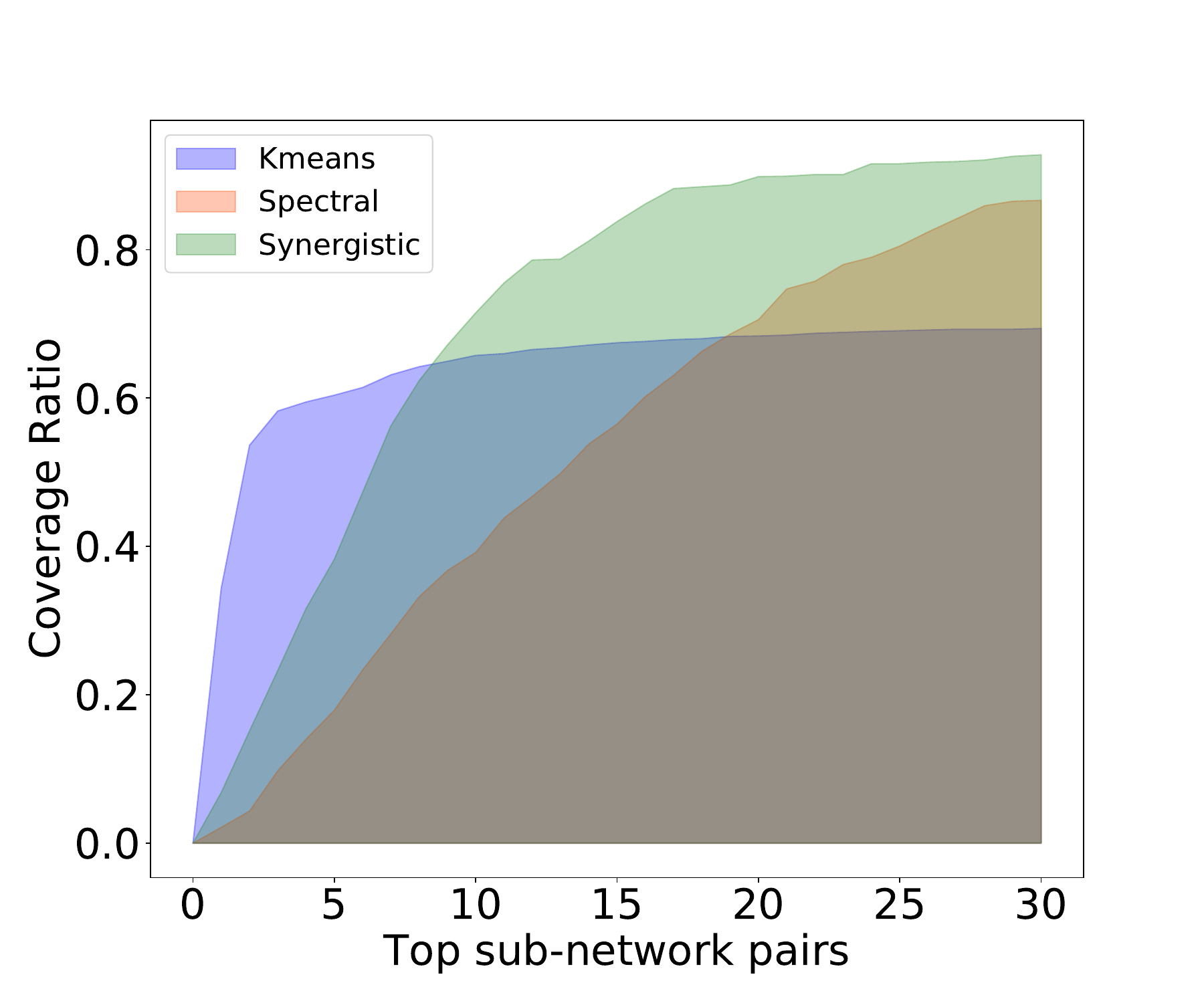}
	\end{minipage}
	\vspace{-10pt}
	\caption{The ratio of covering unknown anchor links}\label{fig:coverage_ratio} 
\end{figure}

\begin{figure}[t]

	\centering
	\begin{minipage}[l]{0.9\columnwidth}
		\centering
		\includegraphics[width=\textwidth]{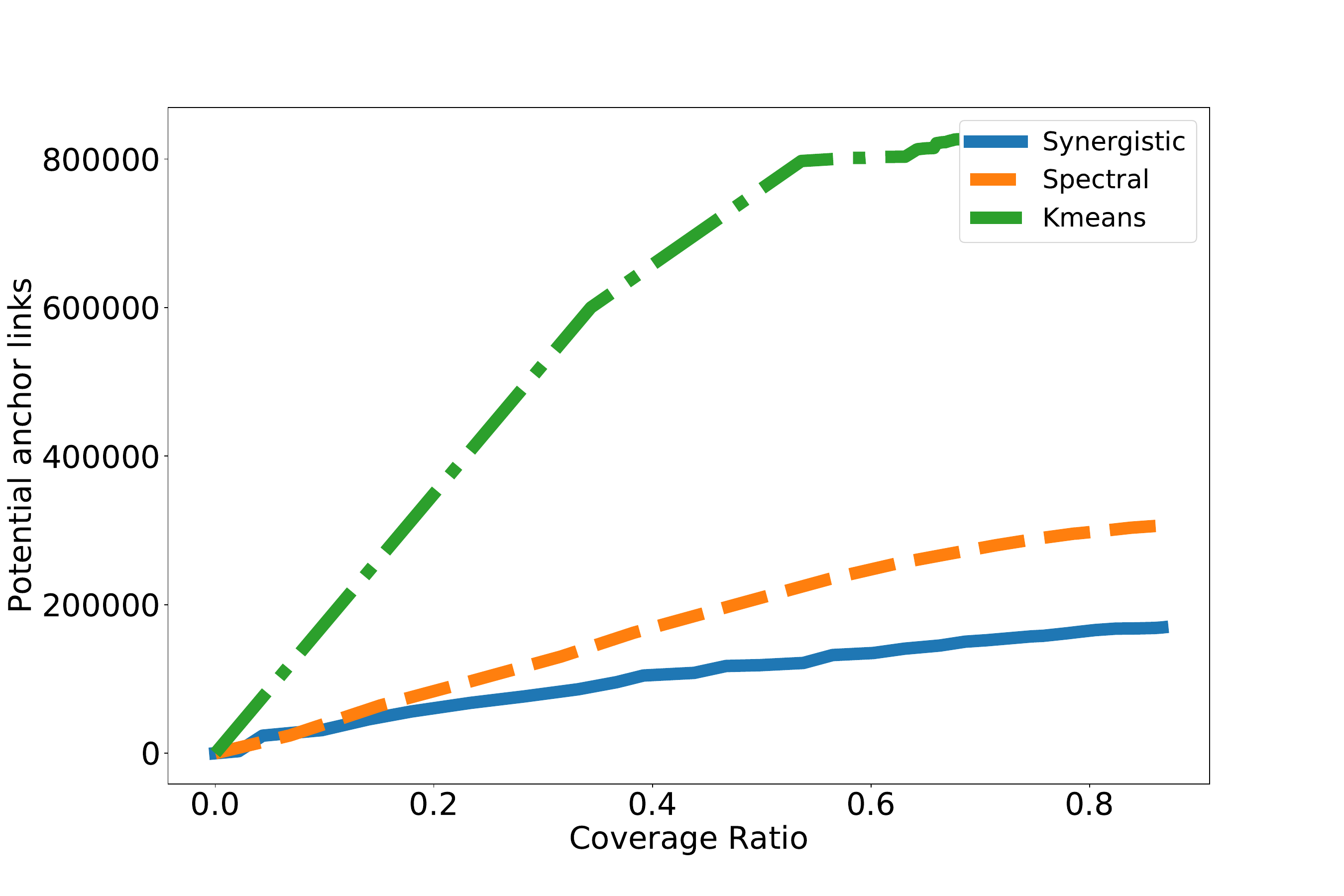}
	\end{minipage}
	\vspace{-10pt}
	\caption{Upward trend of potential anchor links}\label{fig:ratio_total_link} 
\end{figure}

\subsubsection{Partitioned Networks Matching}

Based on the partitioned networks, we can select the optimal sub-network pairs to perform the alignment and reduce the search space by ignoring links not exist in sub-network pairs. We choose the optimal sub-network pairs based on the ranking of M-Score, so to illustrate that our M-Score-based ranking is reliable and effective, we present related experimental results in Figure~\ref{fig:coverage_ratio}. Here, the x-axis denotes the number of selected top sub-network pairs, i.e., $x=5$ means top 5 sub-network pairs in the ranking list are selected for alignment. For the y-axis indicator, we use the truth of the test set where the coverage ratio represents the coverage of positive anchor links in the test set. The reason why the coverage ratio is important is that if the positive anchor links in the test set are not included in selected sub-network pairs, there will be no chance to be predicted to positive in the alignment stage. In other words, positive anchor links are pruned as negative links. What needs to be explained is that the truth of the test set is only used for evaluation here. From Figure~\ref{fig:coverage_ratio} we can find top 30 sub-network pairs from \textit{network synergistic partition} can cover $92.8\%$ positive anchor links which is higher than two other methods with the ratio $0.866$ and $0.693$ respectively. It proves not only the matching policy we used is effective which guarantees that $92.8\%$ positive anchor links have the chance to be predicted, but also the effectiveness of \textit{network synergistic partition}.

From another perspective, the upward trend of the number of potential anchor links with the rise of the coverage ratio can also reflect the performance of negative links pruning. We display the correlation in Figure~\ref{fig:ratio_total_link}, and it is obvious that the increasing rate of \textit{network synergistic partition} is the slowest. It means more impossible and meaningless links are pruned by {\our}, which can affect both time complexity and prediction performance badly in the alignment stage. 
\subsubsection{Parallel Sub-network Alignment}

The experimental results of the alignment stage are shown in Table~\ref{tab:main_result}. 
The methods we test in experiments can all output link prediction labels, and we will use F1, Recall and Precision as evaluation metrics. We will not present the metric Accuracy in the tables, because in such a class-imbalance setting of alignment tasks (the number of negative anchor links is much larger than positive links), the value of Accuracy is not so critical in evaluating the comparison methods.
Firstly, we focus on the comparison among {\our} and {\pusvm}. We can find {\our} has a distinct advantage over {\pusvm} according to all four metrics. It means the alignment task achieves better performance after partition compared with no partition. We insist that \textit{network synergistic partition} not only ensures the scalability but also effectively reduces the search space, that is, pruning impossible and meaningless links. In fact, these links will increase time complexity and affect the alignment stage badly simultaneously. Besides, the comparison among {\our}, {\dw} and {\mv} verifies the effectiveness of inter-network meta diagram based features. It also reminds us that the heterogeneity of social networks needs to be handled in a precise way.
Meanwhile, by comparing {\our}, {\kmeanspu} and {\sppu}, we can demonstrate the partition stage is critical to the alignment stage. {\our} overperforms other methods significantly which verifies the effectiveness of \textit{network synergistic partition} as well. We can observe that the Recall of {\kmeanspu} is lower than {\pusvm} which means the partition based on simple k-means will prune lots of positive anchor links and miss them in the final alignment result.
\begin{figure}[t]
	\centering
	\begin{minipage}[l]{0.9\columnwidth}
		\centering
		\includegraphics[width=\textwidth]{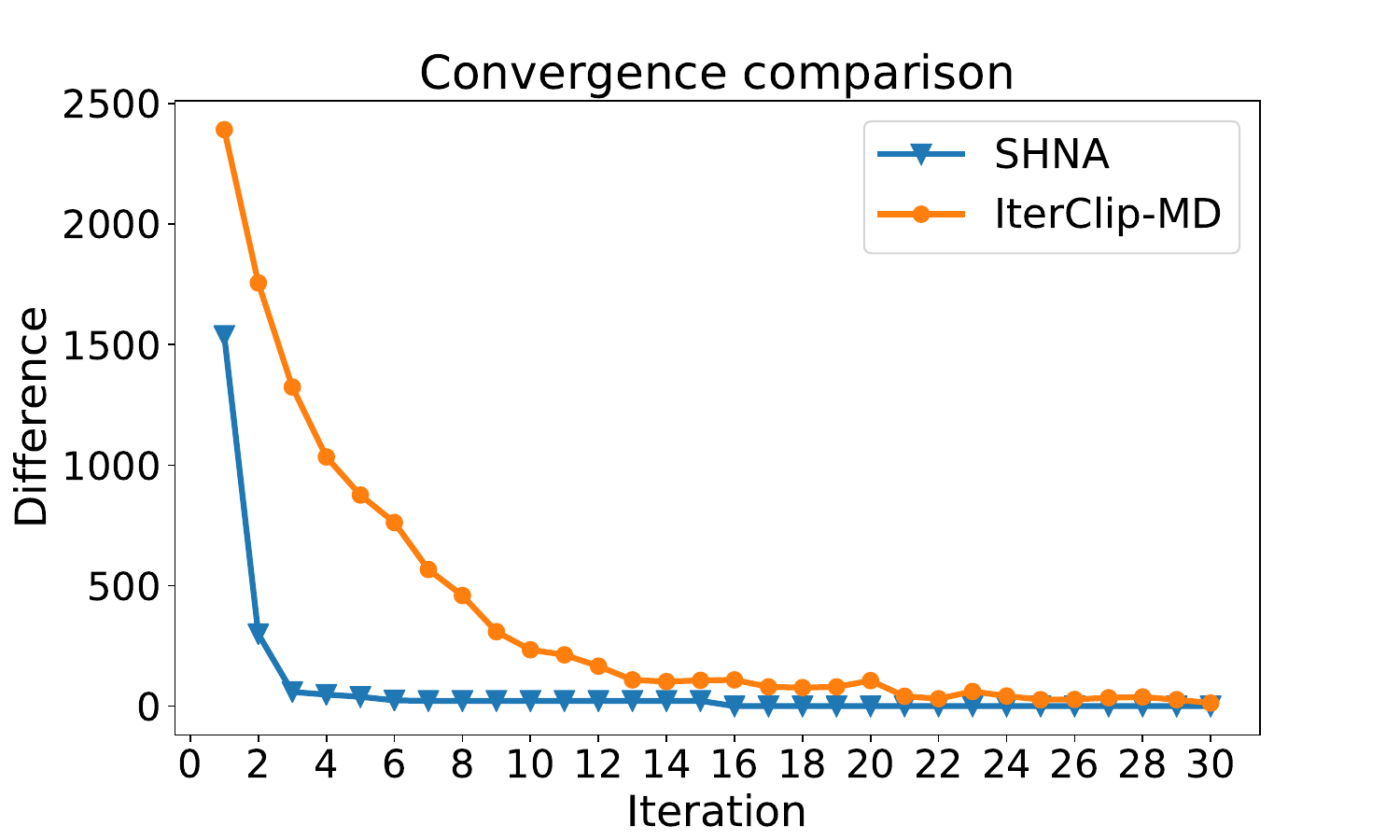}
	\end{minipage}
	\vspace{-5pt}
	\caption{Convergence comparison between {\our} and {\pusvm}}\label{fig:convergence_comparison} 
\end{figure}

\begin{figure}[t]
	\vspace{-10pt}
	\centering
	\begin{minipage}[l]{0.9\columnwidth}
		\centering
		\includegraphics[width=\textwidth]{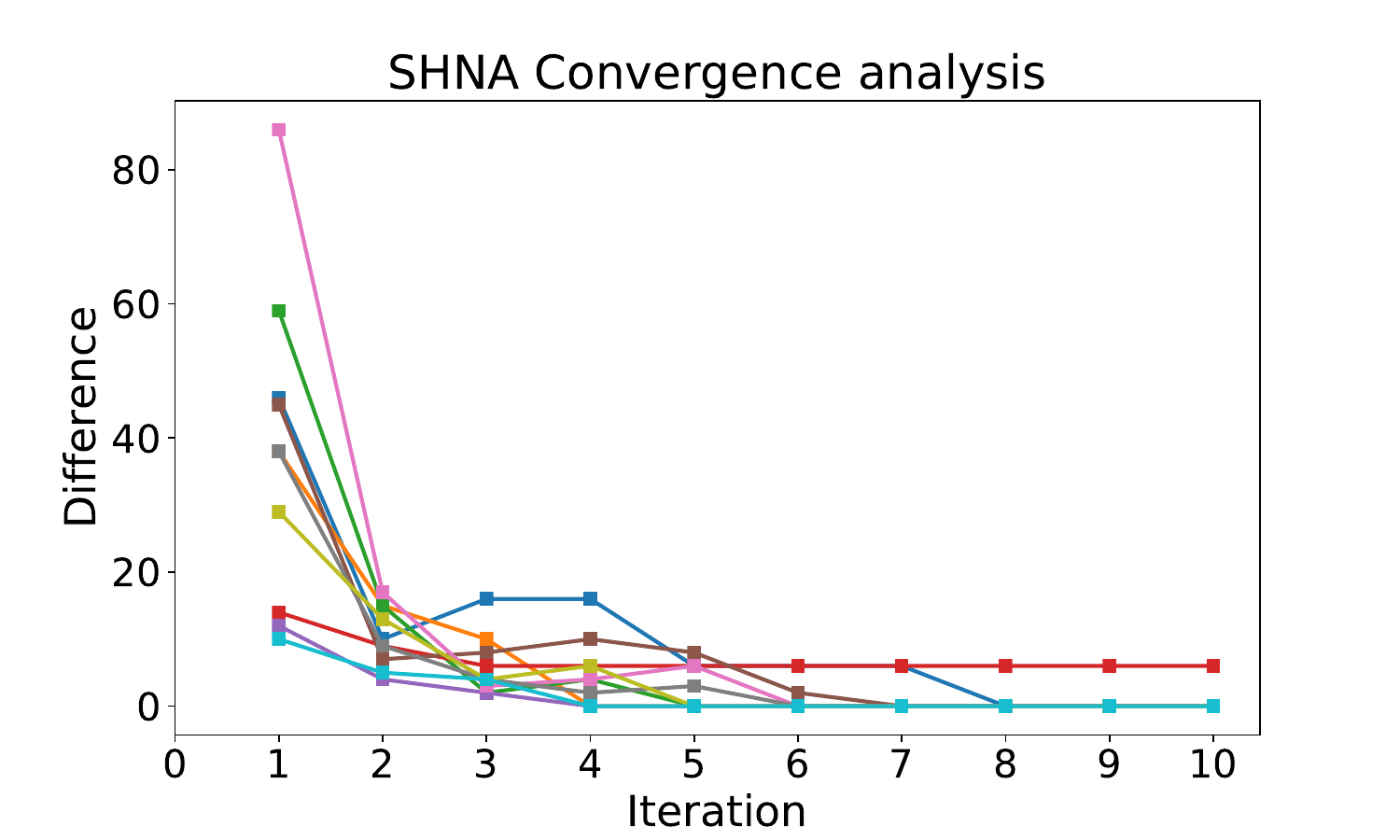}
	\end{minipage}
	\vspace{-1pt}
	\caption{Convergence analysis of {\our}}\label{fig:convergence} 
	\vspace{-20pt}
\end{figure}

\begin{figure*}[t]
	\centering
	\subfigure[F1]{ \label{fig:f1_theta}
		\begin{minipage}[l]{0.47\columnwidth}
			\centering
			\includegraphics[width=1.0\textwidth]{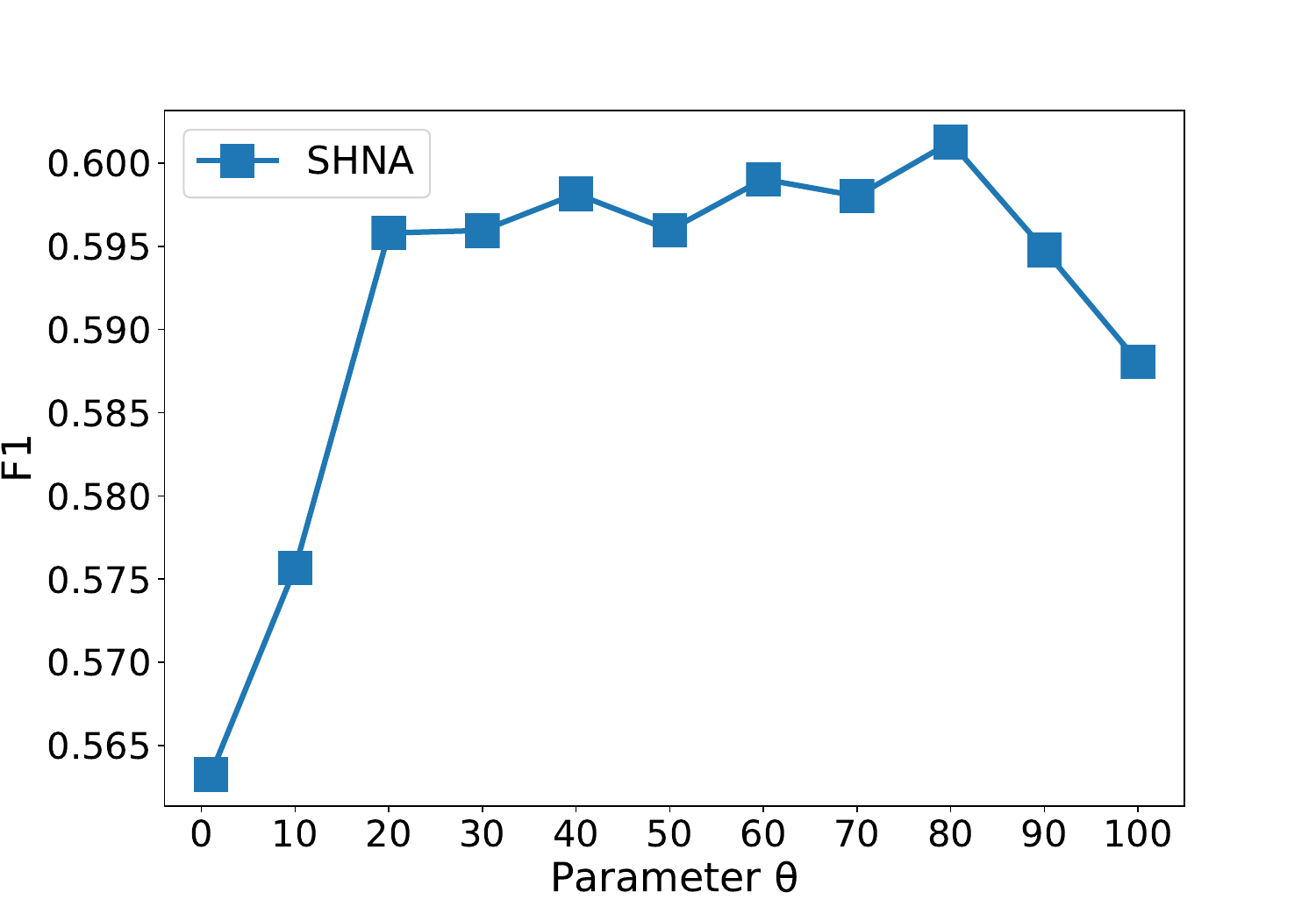}
		\end{minipage}
	}
	\subfigure[Recall]{\label{fig:recall_theta}
		\begin{minipage}[l]{0.47\columnwidth}
			\centering
			\includegraphics[width=1.0\textwidth]{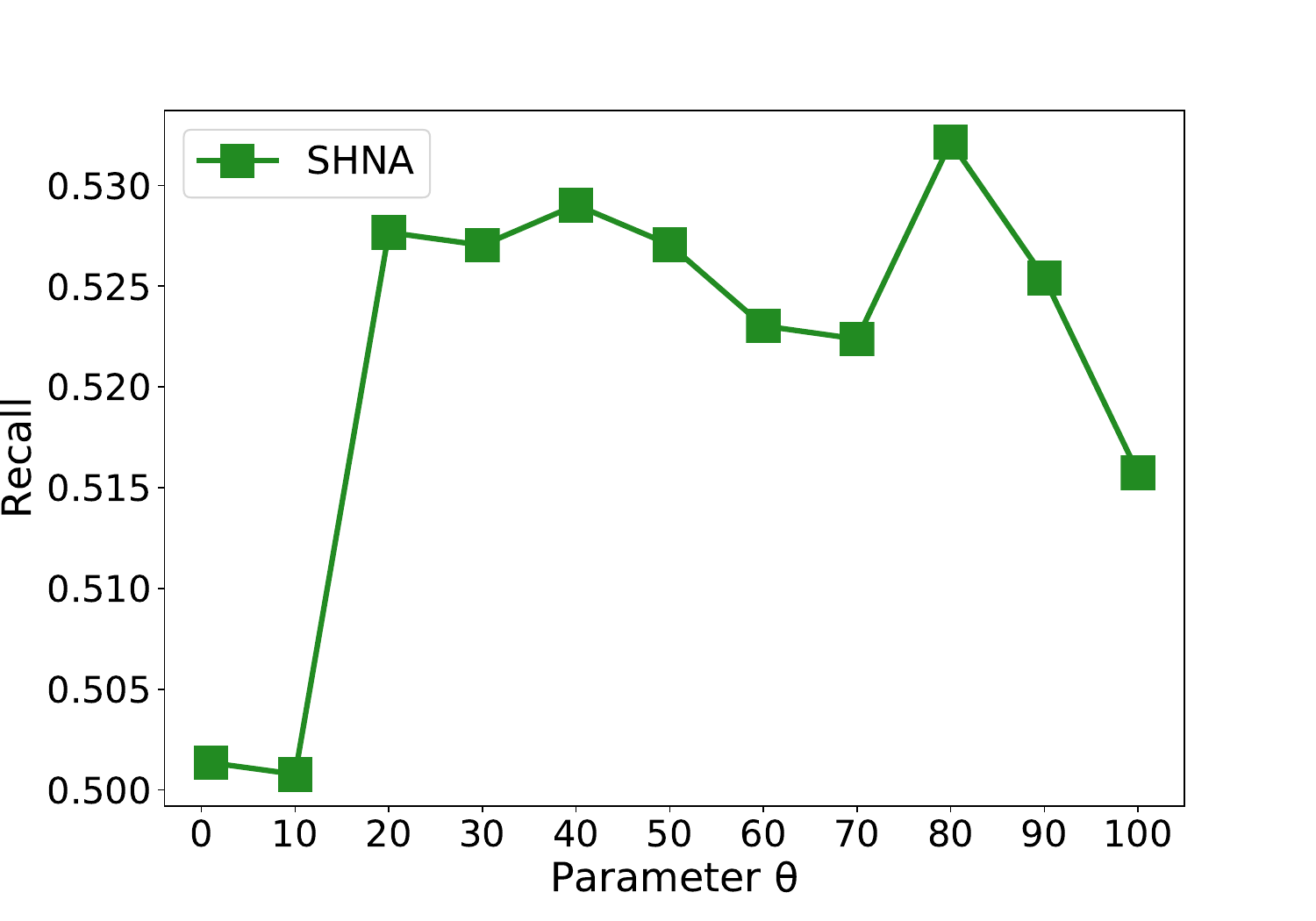}
		\end{minipage}
	}
	\subfigure[Precision]{ \label{fig:precision_theta}
		\begin{minipage}[l]{0.47\columnwidth}
			\centering
			\includegraphics[width=1.0\textwidth]{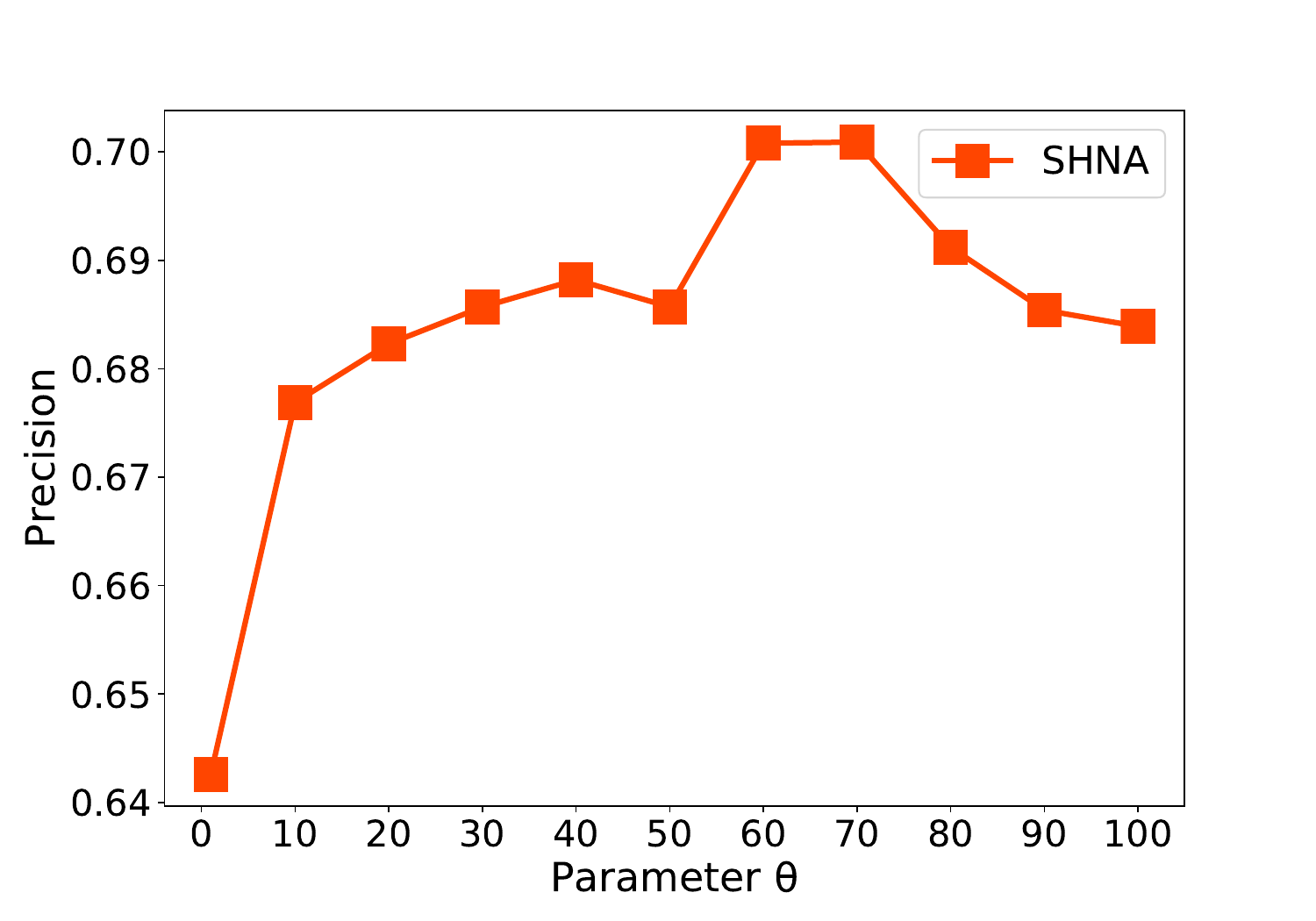}
		\end{minipage}
	}
	\subfigure[Accuracy]{ \label{fig:accuracy_theta}
		\begin{minipage}[l]{0.47\columnwidth}
			\centering
			\includegraphics[width=1.0\textwidth]{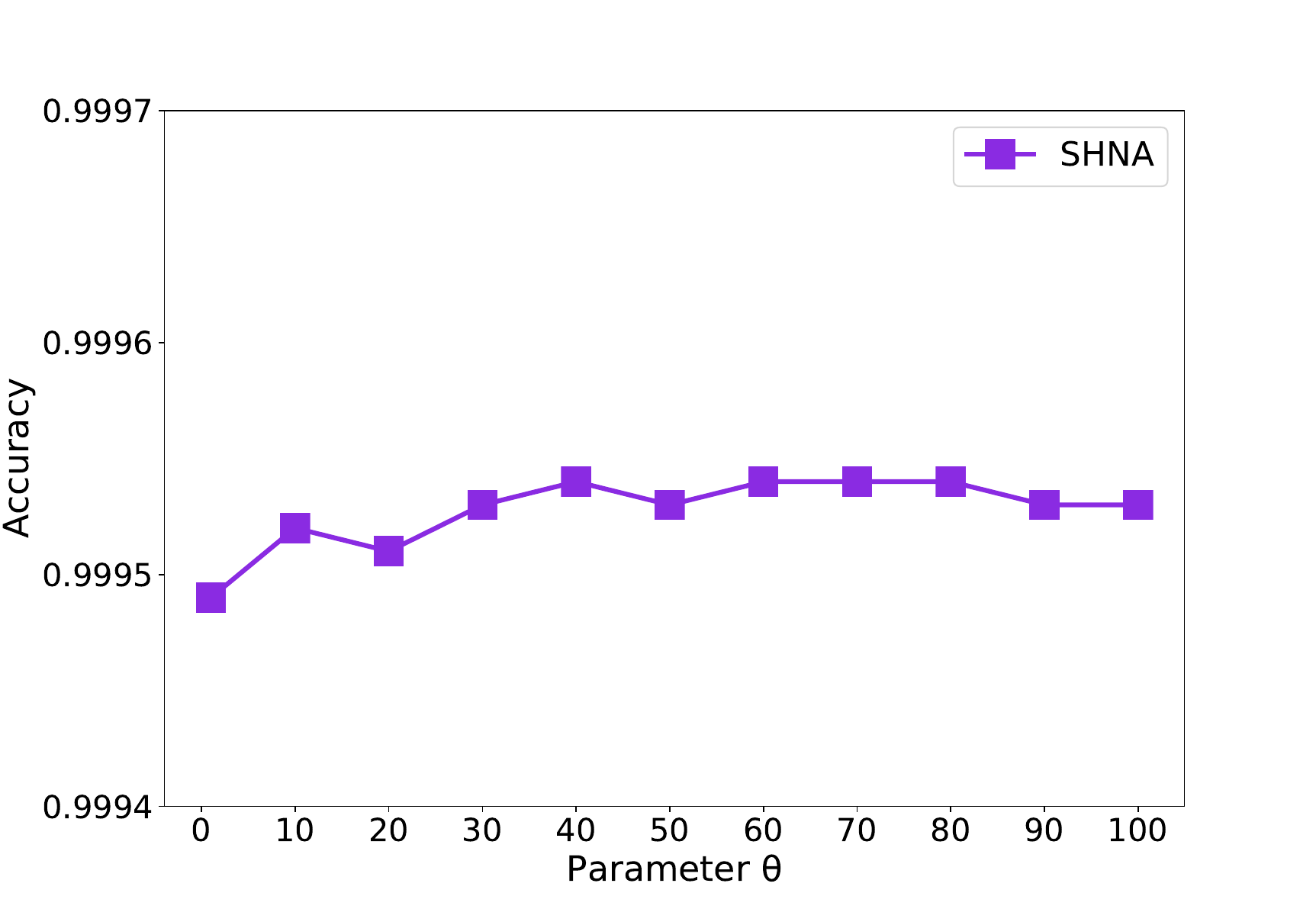}
		\end{minipage}
	}
	\vspace{-5pt}
	\caption{Performance Analysis with different parameter $\theta$}\label{fig:theta_analysis}
	\vspace{-15pt}
\end{figure*}
\subsection{Time and Convergence Analysis}
At first, we compare the convergence between {\our} and {\pusvm}. In building {\our} and {\pusvm}, we propose to use two iteration steps in Section \ref{sec:alignment} to learn the variable vector $\mb{w}$ and predict the anchor link label matrix $\mb{y}$. The number of rounds used to convergence has a significant impact on running time. In Figure~\ref{fig:convergence_comparison}, we show the label matrix $\mb{y}$ changes in each iteration when built {\our} and {\pusvm} respectively. Here, the x axis denotes the iterations, and the y axis denotes the changes of $\mb{y}$ in sequential iterations $i$ and $i-1$, i.e., $\Delta \mb{y} = \left\| \mb{y}^i - \mb{y}^{i-1} \right\|_1$. Because the iteration steps are excuted simultaneously within each subnetwork pair for {\our}, we set the changes of $\mb{y}$ as the sum of the changes of subnetwork pairs, i.e., $\Delta \mb{y} = \sum_{j=1}^{m} \left\| \mb{y}^i_j - \mb{y}^{i-1}_j \right\|_1$. From Figure~\ref{fig:convergence_comparison}, we can find {\our} can reach convergence in much fewer rounds than {\pusvm} where {\pusvm} needs more than 20 rounds to converge, but {\our} converges within 5 rounds. To further illustrate the advantages of {\our} in terms of convergence, we present the convergence speed of top 10 sub-network pairs with the most convergence rounds. Obviously, even top 10 sub-network pairs with the most convergence rounds converge in around 5 rounds. 

The alignment time cost of different methods is listed in Table~\ref{tab:main_result}.
%
{\pusvm} can achieve better prediction results than classic classification methods such as {\svm}, but it costs the longest time. The reason lies in as the size of network increases, the number of rounds required for convergence increases together with each iteration time rises rapidly. {\our} has the best performance in the alignment time cost compared with {\sppu}, {\kmeanspu}, because the partition results of {\sppu}, {\kmeanspu} are uneven and some subnetworks are very large in size. In conclusion, comparison methods with partition can deduct the alignment time cost significantly based on both faster convergence speed and parallel computing. 

\subsection{Parameter Analysis}\label{sec:parameter}

Considering the objective function in Section~\ref{sec:joint-cluster} is composed of $3$ parts, so the weights of different parts are important for the final results. Since the parameters $\alpha$, $\beta$, and $\theta$ mainly reflect the extent to which each part influences the objective function, the proportional relationship among them must be more critical than the numerical values. Therefore, we fix $\alpha = \beta = 1$ because we assume that each network is equally important for partition, and tune $\theta$. From Figure~\ref{fig:discrepancy_theta}, we can observe that the discrepancy keeps monotonous decline with the rising of $\theta$. But combining with Figure~\ref{fig:avg_closeness_theta}, the average M-Score gets the highest value when $\theta=80$ instead of $\theta=100$ which means the value of discrepancy is not the smaller the better for the synergistic partition. We further observe the result of partition with $\theta=1000$ and find that in order to make the discrepancy infinitely close to $0$, most of the anchor links in the training set are concentrated in one pair of sub-networks, and IntraMD-Pro is completely ignored. When $\theta$ is small, the result of partitioning will approximate spectral clustering, because InterMD-Pro will not work due to the insignificant weight. Further, the results of alignment with different $\theta$ can be observed in Figure~\ref{fig:theta_analysis} intuitively. The results show that F1 and Recall obtain the best performance when $\theta=80$. On the contrary, when the value of $\theta$ is too large or too small, the performance becomes worse. In conclusion, $\theta$ should be in a suitable interval to make all parts of the objective function contribute to final results.


\section{Related Work} \label{sec:related_work}
Network alignment has become an important research topic in recent years. Network alignment has concrete applications in various areas, e.g., protein-protein-interaction(PPI) and gene regulatory networks alignment in bioinformatics\cite{KBS08, LLBSB09, SXB07, seah2014dualaligner}, chemical compound matching in chemistry \cite{SHL08}, graph matching in combinatorial mathematics \cite{MH14}, figure matching and merging in computer vision \cite{CFSV04, BGGSW09}, and data schemas matching in data management \cite{MGR02}. Especially in the area of bioinformatics, the network alignment problem aims to predict the optimal mapping between two biological networks.  Network alignment problems can be applied to predict conserved functional modules \cite{SSKKMUSKI05} and to infer protein function \cite{PSBLB11} through exploring the cross-species variation of biological networks. The pairwise network alignment by maximizing the objective function based on a set of learning parameters is proposed by Graemlin \cite{FNSMB06}. The IsoRank proposed in \cite{SXB08} can greedily align multiple networks based on pairwise node similarity scores calculated using spectral theory. IsoRankN \cite{LLBSB09} further extended IsoRank by using a spectral clustering scheme.

In the field of social networks, network alignment provides a powerful tool for information fusion\cite{ZP19} across multiple information sources. Zafarani et al. studies the cross-network user matching problem in \cite{ZAFA13} based on both users relationships and various attribute information. Kong et al.\cite{KZY13} propose to fully align social networks with the heterogeneous link and attribute information simultaneously based on a supervised learning setting. Zhang et al.\cite{ZY15_ijcai, ZCZCY17} propose to study the problem based on the PU learning setting to make use of a small amount known anchor links. A manifold-based method is porposed in  \cite{zhao2018learning} for the social network alignment problem.

Similarity measurement on heterogeneous networks has been widely studied. Sun introduces the concept of \textit{meta path-based similarity} in \cite{SHYYW11}, which can be applied in either link prediction problems \cite{SBGAH11,SHAC12} or clustering problems \cite{SHYYW11,SCH12}. The \textit{meta path} suffers from the disadvantage that cannot describe rich semantics effectively. \textit{Meta structure} \cite{HZCSML16} is proposed to similarity measure problem, but entities are constrained to the same type. Zhao \cite{ZYLSL17} proposes the concept of \textit{meta graph} and extends the idea to recommendation problems. However, \textit{meta structure} and \textit{meta graph} are proposed for the single non-attribute network. 


Clustering-based community detection in online social networks is also related to our {\our} framework. Many different techniques are proposed to optimize certain measures, e.g., modularity function \cite{NG04}, and normalized cut \cite{SM00}. A comprehensive survey of correlated techniques used to detect communities is given by Malliaros et al.\cite{MV13} and a detailed tutorial on spectral clustering is provided by Luxburg \cite{Luxburg07}. These works are mostly studied based on homogeneous networks.
Consensus clustering \cite{MTMG07,LD08,LD13} is a sub-topic under clustering closing to our paper. However, these works mostly aim to find a single consensus clustering from fully mapped clustering solutions. Shao et al.\cite{SZHY16} propose \textit{MMC} which is based on collective spectral clustering with a discrepancy penalty across sources to deal with partially unknown mappings between instances. In comparison, the purpose of partition in our paper is to obtain optimal sub-network matching instead of optimizing the discrepancy merely.

\section{Conclusion}\label{sec:conclusion}
In this paper, we study the heterogeneous social network alignment problem and propose a novel two-stage framework {\our} to solve it. In order to address the extremely large search space, {\our} partitions the original networks with \textit{network synergistic partition}. A group of inter- and intra-network meta diagrams are defined to constitute heterogeneous features. The metrics \textit{Matching Score} is proposed to obtain optimal sub-network matching results. With the support of the partition stage, not only the search space is greatly reduced, but also the alignment within sub-network pairs can be performed in parallel. Extensive experiments are conducted on real-world networks Foursquare and Twitter.
The experiment results demonstrate that {\our} has outstanding performance compared with the state-of-the-art baseline methods in both \textit{network synergistic partition} and \textit{parallel sub-network alignment} stage.

\balance
\bibliographystyle{plain}
\bibliography{reference}

\begin{thebibliography}{10}

\bibitem{BGGSW09}
Mohsen Bayati, Margot Gerritsen, David~F Gleich, Amin Saberi, and Ying Wang.
\newblock Algorithms for large, sparse network alignment problems.
\newblock In {\em 2009 Ninth IEEE International Conference on Data Mining},
  pages 705--710. IEEE, 2009.

\bibitem{CFSV04}
Donatello Conte, Pasquale Foggia, Carlo Sansone, and Mario Vento.
\newblock Thirty years of graph matching in pattern recognition.
\newblock {\em International journal of pattern recognition and artificial
  intelligence}, 18(03):265--298, 2004.

\bibitem{DCS17}
Yuxiao Dong, Nitesh~V Chawla, and Ananthram Swami.
\newblock metapath2vec: Scalable representation learning for heterogeneous
  networks.
\newblock In {\em Proceedings of the 23rd ACM SIGKDD international conference
  on knowledge discovery and data mining}, pages 135--144. ACM, 2017.

\bibitem{du2019joint}
Xingbo Du, Junchi Yan, and Hongyuan Zha.
\newblock Joint link prediction and network alignment via cross-graph
  embedding.
\newblock In {\em Proceedings of the 28th International Joint Conference on
  Artificial Intelligence}, pages 2251--2257. AAAI Press, 2019.

\bibitem{FNSMB06}
Jason Flannick, Antal Novak, Balaji~S Srinivasan, Harley~H McAdams, and Serafim
  Batzoglou.
\newblock Graemlin: general and robust alignment of multiple large interaction
  networks.
\newblock {\em Genome research}, 16(9):1169--1181, 2006.

\bibitem{HZCSML16}
Zhipeng Huang, Yudian Zheng, Reynold Cheng, Yizhou Sun, Nikos Mamoulis, and
  Xiang Li.
\newblock Meta structure: Computing relevance in large heterogeneous
  information networks.
\newblock In {\em Proceedings of the 22nd ACM SIGKDD International Conference
  on Knowledge Discovery and Data Mining}, pages 1595--1604. ACM, 2016.

\bibitem{KBS08}
Maxim Kalaev, Vineet Bafna, and Roded Sharan.
\newblock Fast and accurate alignment of multiple protein networks.
\newblock In {\em Annual International Conference on Research in Computational
  Molecular Biology}, pages 246--256. Springer, 2008.

\bibitem{kazemi2015growing}
Ehsan Kazemi, S~Hamed Hassani, and Matthias Grossglauser.
\newblock Growing a graph matching from a handful of seeds.
\newblock {\em Proceedings of the VLDB Endowment}, 8(10):1010--1021, 2015.

\bibitem{KZY13}
Xiangnan Kong, Jiawei Zhang, and Philip~S Yu.
\newblock Inferring anchor links across multiple heterogeneous social networks.
\newblock In {\em Proceedings of the 22nd ACM international conference on
  Information \& Knowledge Management}, pages 179--188. ACM, 2013.

\bibitem{LD08}
Tao Li and Chris Ding.
\newblock Weighted consensus clustering.
\newblock In {\em Proceedings of the 2008 SIAM International Conference on Data
  Mining}, pages 798--809. SIAM, 2008.

\bibitem{LLBSB09}
Chung-Shou Liao, Kanghao Lu, Michael Baym, Rohit Singh, and Bonnie Berger.
\newblock Isorankn: spectral methods for global alignment of multiple protein
  networks.
\newblock {\em Bioinformatics}, 25(12):i253--i258, 2009.

\bibitem{LD13}
Eric~F Lock and David~B Dunson.
\newblock Bayesian consensus clustering.
\newblock {\em Bioinformatics}, 29(20):2610--2616, 2013.

\bibitem{MV13}
Fragkiskos~D Malliaros and Michalis Vazirgiannis.
\newblock Clustering and community detection in directed networks: A survey.
\newblock {\em Physics Reports}, 533(4):95--142, 2013.

\bibitem{MH14}
Fredrik Manne and Mahantesh Halappanavar.
\newblock New effective multithreaded matching algorithms.
\newblock In {\em 2014 IEEE 28th International Parallel and Distributed
  Processing Symposium}, pages 519--528. IEEE, 2014.

\bibitem{MGR02}
Sergey Melnik, Hector Garcia-Molina, and Erhard Rahm.
\newblock Similarity flooding: A versatile graph matching algorithm and its
  application to schema matching.
\newblock In {\em Proceedings 18th International Conference on Data
  Engineering}, pages 117--128. IEEE, 2002.

\bibitem{MTMG07}
Stefano Monti, Pablo Tamayo, Jill Mesirov, and Todd Golub.
\newblock Consensus clustering: a resampling-based method for class discovery
  and visualization of gene expression microarray data.
\newblock {\em Machine learning}, 52(1-2):91--118, 2003.

\bibitem{NG04}
Mark~EJ Newman and Michelle Girvan.
\newblock Finding and evaluating community structure in networks.
\newblock {\em Physical review E}, 69(2):026113, 2004.

\bibitem{PSBLB11}
Daniel Park, Rohit Singh, Michael Baym, Chung-Shou Liao, and Bonnie Berger.
\newblock Isobase: a database of functionally related proteins across ppi
  networks.
\newblock {\em Nucleic acids research}, 39(suppl\_1):D295--D300, 2010.

\bibitem{PAS14}
Bryan Perozzi, Rami Al-Rfou, and Steven Skiena.
\newblock Deepwalk: Online learning of social representations.
\newblock In {\em Proceedings of the 20th ACM SIGKDD international conference
  on Knowledge discovery and data mining}, pages 701--710. ACM, 2014.

\bibitem{RAZ19}
Yuxiang Ren, Charu~C Aggarwal, and Jiawei Zhang.
\newblock Meta diagram based active social networks alignment.
\newblock In {\em 2019 IEEE 35th International Conference on Data Engineering
  (ICDE)}, pages 1690--1693. IEEE, 2019.

\bibitem{seah2014dualaligner}
Boon-Siew Seah, Sourav~S Bhowmick, and C~Forbes Dewey~Jr.
\newblock Dualaligner: a dual alignment-based strategy to align protein
  interaction networks.
\newblock {\em Bioinformatics}, 30(18):2619--2626, 2014.

\bibitem{SZHY16}
Weixiang Shao, Jiawei Zhang, Lifang He, and S~Yu Philip.
\newblock Multi-source multi-view clustering via discrepancy penalty.
\newblock In {\em 2016 International Joint Conference on Neural Networks
  (IJCNN)}, pages 2714--2721. IEEE, 2016.

\bibitem{SSKKMUSKI05}
Roded Sharan, Silpa Suthram, Ryan~M Kelley, Tanja Kuhn, Scott McCuine, Peter
  Uetz, Taylor Sittler, Richard~M Karp, and Trey Ideker.
\newblock Conserved patterns of protein interaction in multiple species.
\newblock {\em Proceedings of the National Academy of Sciences},
  102(6):1974--1979, 2005.

\bibitem{SM00}
Jianbo Shi and Jitendra Malik.
\newblock Normalized cuts and image segmentation.
\newblock {\em Departmental Papers (CIS)}, page 107, 2000.

\bibitem{SXB07}
Rohit Singh, Jinbo Xu, and Bonnie Berger.
\newblock Pairwise global alignment of protein interaction networks by matching
  neighborhood topology.
\newblock In {\em Annual International Conference on Research in Computational
  Molecular Biology}, pages 16--31. Springer, 2007.

\bibitem{SXB08}
Rohit Singh, Jinbo Xu, and Bonnie Berger.
\newblock Global alignment of multiple protein interaction networks with
  application to functional orthology detection.
\newblock {\em Proceedings of the National Academy of Sciences},
  105(35):12763--12768, 2008.

\bibitem{SHL08}
Aaron Smalter, Jun Huan, and Gerald Lushington.
\newblock Gpm: A graph pattern matching kernel with diffusion for chemical
  compound classification.
\newblock In {\em 2008 8th IEEE International Conference on BioInformatics and
  BioEngineering}, pages 1--6. IEEE, 2008.

\bibitem{SCH12}
Yizhou Sun, Charu~C Aggarwal, and Jiawei Han.
\newblock Relation strength-aware clustering of heterogeneous information
  networks with incomplete attributes.
\newblock {\em Proceedings of the VLDB Endowment}, 5(5):394--405, 2012.

\bibitem{SBGAH11}
Yizhou Sun, Rick Barber, Manish Gupta, Charu~C Aggarwal, and Jiawei Han.
\newblock Co-author relationship prediction in heterogeneous bibliographic
  networks.
\newblock In {\em 2011 International Conference on Advances in Social Networks
  Analysis and Mining}, pages 121--128. IEEE, 2011.

\bibitem{SHAC12}
Yizhou Sun, Jiawei Han, Charu~C Aggarwal, and Nitesh~V Chawla.
\newblock When will it happen?: relationship prediction in heterogeneous
  information networks.
\newblock In {\em Proceedings of the fifth ACM international conference on Web
  search and data mining}, pages 663--672. ACM, 2012.

\bibitem{SHYYW11}
Yizhou Sun, Jiawei Han, Xifeng Yan, Philip~S Yu, and Tianyi Wu.
\newblock Pathsim: Meta path-based top-k similarity search in heterogeneous
  information networks.
\newblock {\em Proceedings of the VLDB Endowment}, 4(11):992--1003, 2011.

\bibitem{Luxburg07}
Ulrike Von~Luxburg.
\newblock A tutorial on spectral clustering.
\newblock {\em Statistics and computing}, 17(4):395--416, 2007.

\bibitem{wang2018deepmatching}
Chenxu Wang, Zhiyuan Zhao, Yang Wang, Dong Qin, Xiapu Luo, and Tao Qin.
\newblock Deepmatching: A structural seed identification framework for social
  network alignment.
\newblock In {\em 2018 IEEE 38th International Conference on Distributed
  Computing Systems (ICDCS)}, pages 600--610. IEEE, 2018.

\bibitem{ZAFA13}
Reza Zafarani and Huan Liu.
\newblock Connecting users across social media sites: a behavioral-modeling
  approach.
\newblock In {\em Proceedings of the 19th ACM SIGKDD international conference
  on Knowledge discovery and data mining}, pages 41--49. ACM, 2013.

\bibitem{ZCZCY17}
Jiawei Zhang, Jianhui Chen, Junxing Zhu, Yi~Chang, and Philip~S Yu.
\newblock Link prediction with cardinality constraint.
\newblock In {\em Proceedings of the Tenth ACM International Conference on Web
  Search and Data Mining}, pages 121--130. ACM, 2017.

\bibitem{ZY15_ijcai}
Jiawei Zhang and S~Yu Philip.
\newblock Integrated anchor and social link predictions across social networks.
\newblock In {\em Twenty-Fourth International Joint Conference on Artificial
  Intelligence}, 2015.

\bibitem{ZY15-2}
Jiawei Zhang and Philip~S Yu.
\newblock Community detection for emerging networks.
\newblock In {\em Proceedings of the 2015 SIAM International Conference on Data
  Mining}, pages 127--135. SIAM, 2015.

\bibitem{ZP19}
Jiawei Zhang and Philip~S Yu.
\newblock {\em Broad Learning Through Fusions: Applications in Machine
  Learning}.
\newblock Springer, 2019.

\bibitem{zhang2018mego2vec}
Jing Zhang, Bo~Chen, Xianming Wang, Hong Chen, Cuiping Li, Fengmei Jin, Guojie
  Song, and Yutao Zhang.
\newblock Mego2vec: embedding matched ego networks for user alignment across
  social networks.
\newblock In {\em Proceedings of the 27th ACM International Conference on
  Information and Knowledge Management}, pages 327--336. ACM, 2018.

\bibitem{ZYLSL17}
Huan Zhao, Quanming Yao, Jianda Li, Yangqiu Song, and Dik~Lun Lee.
\newblock Meta-graph based recommendation fusion over heterogeneous information
  networks.
\newblock In {\em Proceedings of the 23rd ACM SIGKDD International Conference
  on Knowledge Discovery and Data Mining}, pages 635--644. ACM, 2017.

\bibitem{zhao2018learning}
Wei Zhao, Shulong Tan, Ziyu Guan, Boxuan Zhang, Maoguo Gong, Zhengwen Cao, and
  Quan Wang.
\newblock Learning to map social network users by unified manifold alignment on
  hypergraph.
\newblock {\em IEEE transactions on neural networks and learning systems},
  29(12):5834--5846, 2018.

\end{thebibliography}

\end{document}